\documentclass[showpacs,aps,prd,reprint,groupedaddress,superscriptaddress,preprintnumbers,floatfix,a4paper,nofootinbib,amsmath,amssymb]{revtex4-1}

\usepackage{color}
\usepackage{amsmath}
\usepackage{footmisc}
\usepackage{amssymb}
\usepackage{dcolumn}  
\usepackage{slashed}
\usepackage{isotope}
\usepackage{graphicx}
\usepackage{mathtools}
\usepackage[utf8]{inputenc}
\usepackage{multirow}
\usepackage[shortlabels]{enumitem}
\usepackage{hyperref}
\usepackage{soul}
\usepackage{dutchcal}
\hypersetup{colorlinks, linkcolor = [rgb]{0,0.0,0.75}, citecolor = [rgb]{0,0.0,0.75}, urlcolor = [rgb]{0,0.0,0.75}}

\makeatletter
\g@addto@macro\bfseries{\boldmath}
\makeatother

\begin{document}

\preprint{TIFR/TH/24-20}

\title{Spectrum of two-flavored spin-zero heavy dibaryons in lattice QCD}

\author{Parikshit M. Junnarkar}
\email{parry@rcnp.osaka-u.ac.jp}
\affiliation{Research Centre for Nuclear Physics, Osaka University \\ 10-1 Mihogaoka, Ibaraki 567-0047, Japan.}
\affiliation{Interdisciplinary Theoretical and Mathematical Sciences Program (iTHEMS), RIKEN, Wako 351-0198, Japan}

\author{Nilmani Mathur}
\email{nilmani@theory.tifr.res.in}
\affiliation{Department of Theoretical Physics,\\
Tata Institute of Fundamental Research,
1 Homi Bhabha Road, Mumbai 400005, India.}

\begin{abstract}
	We present the ground state energy spectra of two-flavored heavy dibaryons in the spin-singlet channel. 
	In particular, the ground state masses of $\Omega_{llQ}\Omega_{llQ}, \Omega_{QlQ}\Omega_{llQ}$ and $\Omega_{QlQ}\Omega_{QlQ}$ states are computed and compared with their respective lowest non-interacting energy levels, where the flavor $Q \in (c,b)$ denotes charm  and bottom quarks, and the other flavor $l \in {s,c,b}$ represents strange, charm and bottom, respectively. Considering their valence quark structures,  these hadrons could be thought of as the heavy flavor analogues of spin-singlet nucleon-nucleon states. 
	The gauge configurations employed in this study are  HISQ ensembles with $N_f = 2+1+1$ flavors, generated by the MILC collaboration, at four lattice spacings, namely $a=0.1207, 0.0888, 0.0582$ and $0.0448$ fm. 
	The aforementioned states are also computed at different quark masses, between $m_s \le m_l \le m_b$, including at unphysical heavy quark masses,  to explore the quark mass dependence of any possible binding.
For the dibaryon states
$\Omega_{bbc}\Omega_{bbc}, \Omega_{ccb}\Omega_{ccb}$ and $\Omega_{ccb} \Omega_{bbc}$,
we find a clear evidence of an energy level below their respective non-interacting energy levels. 
In addition, for these dibaryons at quark masses  $m_c < m_l \le m_b$, a trend is found where the gap, between the lowest energy levels and the respective lowest non-interacting levels, increases as the quark mass $m_l$ increases.
We also study the heavy quark spin-symmetry and its breaking for these heavy dibaryons.
\end{abstract}


\maketitle
\section{Introduction}
Quantum chromodynamics (QCD), the theory of strong interactions, predicts a rich spectrum of hadronic states with quarks and gluons.
Historically, within the quark models, these states have been described quite successfully  in terms of valence quark content in the form of mesons (quark-antiquark) and baryons (three quark).
However, QCD presents no inherent restriction in the formation of other color neutral states, for example, of the form of only gluons (glueballs), a combination of quarks and gluons (hybrids) and multiquark configurations in the form of tetraquark (two quarks together with two antiquarks), pentaquark (four quarks with an anti-quark) and hexaquark/dibaryon (six quarks) hadrons.
All such states, beyond those of the quark model, are generally also called exotic hadrons.
The physics of exotic states is expected to be different than those of the usual mesons and baryons.
Although there are many predictions of exotic states in the light quark sector (with flavors $u,d,s$), the experimental search for them is rather quite non-trivial \cite{olsen2018nonstandard}, due to large mixing between various possible hadronic configurations with numerous decay modes and consequent broad widths. In contrast, exotic states with heavy quarks (bottom $b$, and charm $c$) may exhibit less mixing, and some of them may have only weak decay modes, which makes their identification and discovery more feasible in experiments. Indeed in the past decade and a half, there have been discoveries of many exotic hadrons with at least one heavy quark content in their valence structures. For example, hadrons with fourquark contents, such as $T_{b\bar{b}1}$ \cite{Belle:2011aa}, $T_{c\bar{c}1}$\cite{BESIII:2017bua}, $T_{c\bar{c}\bar{s}1}$ \cite{LHCb:2021uow}\footnote{The notation here for tetraquarks is adopted from Particle Data Group (PDG)~\cite{PDG:2024cfk}}, $T_{\bar{c}c}$~\cite{BESIII:2015tix},
and more recently the $T_{cc}$ state \cite{LHCb:2021vvq}, and in the pentaquarks,
such as $P_{c\bar{c}}(4312)$ and $P_{c\bar{c}}(4450)$ \cite{LHCb:2019kea} have been discovered. See the reviews in Refs. 
\cite{ex-review,Richard:2016eis,BRAMBILLA20201,LIU2019237} and a compilation on exotic states in Ref.~\cite{QWGEH}
for more details.
It is thus natural to investigate 
similar states also with six-quark contents, with one or more heavy quarks in their valence structures,  and look for their discovery as and when required center-of-momentum energy and luminosity become available at the experimental facilities. 

There have been several theoretical studies on  dibaryons with heavy quarks in recent years.
These studies have been directed towards heavy flavor analogues of the $H$-dibaryons through constituent quark model, one pion exchange model  \cite{Carames:2015sya,Richard:2016eis,Meguro:2011nr}.
Heavy quark symmetries such as the heavy-antiquark-diquark symmetry has applied to dibaryons establishing a relation to pentaquarks such as Refs.\cite{Pan:2019skd,PhysRevD.102.011504}.
In addition, two flavored dibaryons have also been studied using quark model in Ref.~\cite{Richard:2020zxb} as well as single flavor in Ref.~\cite{Weng:2022ohh} in a chromomagnetic model which includes color electric and  magnetic interactions.
However, all these model-based calculations, despite their impressive results, may not be fully suitable for capturing the non-perturbative aspects of these states, particularly the quark mass dependence of any binding that may lead to identifying bound dibaryons with  heavy quarks. For that it is important to investigate these states with non-perturbative first-principles method, and Lattice QCD provides such a framework.

Lattice QCD is a first principles method for calculating observables directly from QCD with controlled approximations, which can be improved systematically.
This method is therefore well suited for calculating hadronic observables and has been used to great effect for single hadrons.
In recent years, research in Lattice QCD has also been intensified in the exploration of dibaryon systems both with light quarks \cite{Detmold:2024iwz,Orginos:2015aya,Beane:2012ey,NPLQCD:2011naw,Buchoff:2012ja,Berkowitz:2015eaa,Inoue:2010es,Gongyo:2020pyy,Inoue:2011ai,Gongyo:2017fjb} as well as with heavy quarks \cite{Junnarkar:2019equ,Junnarkar:2022yak,Mathur:2022ovu,Lyu:2021qsh,Miyamoto:2017tjs}.
There are several challenges in calculating multi-hadron correlation functions with Lattice QCD. 
One of the them  is the large number of Wick contractions that arises due to presence of 12 quarks in a source-sink setup. However, in recent years several methods have been developed to to mitigate the cost of Wick-contractions as in Refs. ~\cite{Detmold:2012eu,Doi:2012xd,Humphrey:2022yjc}.
Another important issue is that of the spectroscopy of multi-hadron states, where a reliable extraction of ground state energy level is crucial. In that, the method of distillation has been shown to be quite effective ~\cite{Morningstar:2011ka}.
Finally, to determine the pole structure of a possible state, one needs to perform scattering amplitudes analysis \cite{Luscher:1990ck} using finite volume spectrum obtained at different physical volumes and different momentum frames.
All of these challenges make a lattice calculation of multi-hadron states a numerically intensive effort.
For the specific case of dibaryon with heavy quarks, there are some advantages associated with heavy quarks namely that of better signal-to-noise ratio in correlation functions since a heavy quark propagator is less prone to fluctuations due to its large mass. This leads  to relatively more pronounce plateaus in their ground state effective mass in comparison to those of light hadrons. Another advantage with heavy dibaryons comes from the suppression of finite volume effects, especially in the case of bound states~\cite{Junnarkar:2018twb,Junnarkar:2019equ,Junnarkar:2022yak}.

Despite the above-mentioned challenges, Lattice QCD-based calculations of dibaryons have progressed over the past two decades. 
In the light quark sector, the octet-octet dibaryon channels have been investigated for more than a decade both with the Lüscher method~\cite{Detmold:2024iwz,Orginos:2015aya,Beane:2012ey,NPLQCD:2011naw,Francis:2016hui,Buchoff:2012ja,Berkowitz:2015eaa} as well as with the HALQCD method~\cite{Nemura:2017vjc,Inoue:2010es,Gongyo:2020pyy,Inoue:2011ai,Gongyo:2017fjb}.
Two particular candidates namely the $H$-dibaryon~\cite{NPLQCD:2010ocs,Inoue:2010es,Luo:2011ar,Francis:2018qch,Green:2021qol} and the deuteron~\cite{NPLQCD:2011naw,Doi:2017cfx} have been the focused of such investigations due to their phenomenological impact in understanding nuclear bindings. In the heavy quark sector, there have been a few dibaryon calculations with one or more heavy quarks in recent years~\cite{Junnarkar:2019equ,Junnarkar:2022yak,Mathur:2022ovu,Lyu:2021qsh}.
These calculations with heavy quarks can be classified in general according to their number of flavors, and the main goal of these studies have been to identify the  presence of bound states, if any, below their respective non-interacting energy thresholds. 
There have been two single flavor heavy quark calculations both with  charm and bottom quarks. 
The spin-3/2 $\Omega_{ccc} \Omega_{ccc}$ state has been explored by the HALQCD~\cite{Lyu:2021qsh} where a shallow bound state was found, and with the inclusion of Coulomb interaction a dibaryon system near the unitary limit was found.
In the bottom sector, the interaction of two spin-3/2 $\Omega_{bbb}$ yielded a deeply bound state, and the inclusion of Coulomb effects induces a decrease of $5-10$ MeV in the binding energy~\cite{Mathur:2022ovu}.
With two flavors, the work in Ref.~\cite{Junnarkar:2019equ}, explored the deuteron quantum numbers with different heavy flavor combinations searching for energy levels below the non-interacting levels. 
The  charm-bottom sector yielded the most encouraging results and in addition the binding energy was shown to become stronger with the increase of heavy quark mass.
With three flavors, the dibaryon calculation in Ref.~\cite{Junnarkar:2022yak} 
extended the $H$-dibaryon quantum numbers to heavy flavors and explored the ground state of various flavor combinations.
The combination with bottom-charm-strange was found to be a potentially attractive channel for the presence of a bound state.

In this work we extend 
the lattice QCD study of heavy dibaryons to the two-flavored spin-zero sector. This is similar to 
Ref.~\cite{Junnarkar:2019equ}
where a similar study but in spin-one sector was conducted.
The importance of the exploration of nucleon-nucleon ($NN$) interactions and its implications to nuclear physics can hardly be overstated.
In this sense, the current study is an extension of the $NN$ interactions in the spin-singlet channel where the dineutron belongs, but with heavy flavors.
The extension to heavy quarks naturally has a large breaking of the isospin symmetry and hence the corresponding states, which are almost degenerate with light quarks, will be distinct since there is no flavor symmetry between them. 
A further motivation for this study comes from the results of the work in Ref.~\cite{Junnarkar:2019equ}, which provided encouraging observation in the bottom-charm sector for the possible existence of a bound  states.
Similar to that work ~\cite{Junnarkar:2019equ}, there is also an opportunity to study the heavy quark mass dependence of these states. 
Additionally, with the dynamics of  heavy quarks, it would be natural to contemplate on emergence of heavy quark symmetries in these dibaryons.
As such, here we explore the flavor space in the spin-singlet channel and compute the ground state correlation function of two spin-1/2 baryons, such as $\Omega_{llQ} \Omega_{llQ}, \Omega_{llQ} \Omega_{QlQ}$ and $\Omega_{QlQ} \Omega_{QlQ}$, where the flavors $(l,Q) \in (s,c,b)$ are strange, charm and bottom respectively. 

The contents of the paper are summarized here. 
In section \ref{sec:latt}, the lattice setup used in this work is described where details of the valence and sea actions are provided, along with the heavy quark actions. Next we discuss
the various dibaryon operators considered in this work.
In section \ref{sec:results}, the results of ground state spectrum extracted from various dibaryon operators are presented starting with the pattern of non-interacting (NI) energy levels. Then we present the results in the bottom and charm sectors followed by the heavy quark mass dependence of the effective binding energies. A discussion is followed thereafter on the heavy quark spin symmetry between the spin-zero and spin-1 dibaryons
In section \ref{sec:disc}, a discussion of the results and the conclusions from this work are presented.
 
\section{Lattice set-up and operators \label{sec:latt}}
The lattice setup that we employ in this work is similar to the one utilized in previous works, namely in Refs. \cite{Junnarkar:2018twb,Junnarkar:2019equ,Junnarkar:2022yak}. For completeness we summarize it here briefly. 
We use four $N_f = 2 + 1 + 1$ lattice QCD ensembles each at a different lattice spacing generated by the MILC collaboration~\cite{Bazavov:2012xda}. The dynamical quark flavors are realized with Highly Improved Staggered
Quark (HISQ) action on gauge fields that respect one loop, tadpole-improved Symanzik gauge action. The sea strange and charm quark masses  tuned to their physical values. 
The parameters of these are listed in Table \ref{tab:pars}.
\begingroup
\renewcommand*{\arraystretch}{1.5}
\begin{table}[ht]
	\centering
	\begin{tabular}{cccc} \hline \hline 
	$L^3 \times T$     & $m^{\text{sea}}_\pi$ (MeV) & $m_\pi \text{L}$ & $a$ (fm)\\ \hline \hline    
	$24^3 \times 64 $  &              305.3           & 4.54       & 0.1207(11) \\ \hline 
    $32^3 \times 96 $  &              312.7           & 4.50       & 0.0888(8) \\ \hline 
	$48^3 \times 144 $ &              319.3           & 4.51       & 0.0582(5) \\ \hline 
	$64^3 \times 192 $ &              319.3           & 4.51       & 0.0448(5) \\ \hline \hline
	\end{tabular}
	\caption{\label{tab:pars}{Parameters of lattice QCD ensembles used in this work.}}
	\end{table} 
\endgroup
The lattice spacing was determined using the $\Omega$ baryon and is found to be consistent with that obtained with the $r_1$ parameter by the MILC collaboration~\cite{Bazavov:2015yea} using Wilson flow.

In the valence sector, a relativistic overlap action is used for light to charm flavors. The overlap action does not have ${\cal{O}}(ma)$ error which helps in controlling the discretization error for heavy quarks. Recently it has been shown that the reliable extraction of dibaryon energy levels is crucially dependent on discretization effects \cite{Green:2021qol}. In this sense the use of overlap action is beneficial.
The strange quark mass is tuned by setting the mass of unphysical pseudoscalar  $\bar{s}s$ to 688 MeV \cite{PhysRevD.91.054508}.
The charm quark mass is tuned by setting the spin-averaged charmonia ${1\over 4}(3M_{J/\psi}+M_{\eta_c})$ kinetic mass to its experimental value.
Quark propagators are computed with a gauge-fixed wall source at several quark masses at each lattice spacing.
The corresponding pseudoscalar meson masses are listed in Table \ref{tab:mqs}.
\begingroup
\renewcommand*{\arraystretch}{1.2}
\begin{table}[ht]
	\centering
	\begin{tabular}{ccc}\hline \hline
		$L^3 \times T$  & $a$ (fm) & $m_{ps}$ (MeV)  \\ \hline \hline
		$24^3 \times 64$    & 0.1207(11)    &  688 \\ 
	\hline \hline
		$32^3 \times 96$ & 0.0888(8)&  688   \\ 
            \hline \hline                                                                                $48^3 \times 144$ & 0.0582(5)  & 9399\\
                               &            & 6175\\
                               &            & 5146 \\
                               &            & 4120 \\                                                          &            & 2984  \\           
                               &            & 688  \\
	
		 \hline \hline	
           $64^3 \times 192$ & 0.0448(5) & 688 \\
                             &  & 2984 \\
                             \hline \hline
   \end{tabular}
	\caption{\label{tab:mqs}{Range of pseudoscalar meson masses ($m_{ps}$) corresponding to a number of bare quark masses that we used in this work on each of the ensembles. }}
\end{table} 
\endgroup	
For the bottom flavor, a NRQCD action is employed where the Hamiltonian comprises of terms up to $1/(am_b)^2$ and leading order term of $1/(am_b)^3$, with $m_b$ as the bare bottom quark mass.
The NRQCD Hamiltonian is written as $H = H_0 + \Delta H$, where  the interaction term, $\Delta H$, as used here, is given by,
\begin{eqnarray}
\Delta H &=& -c_1 \frac{(\Delta^{(2)})^2}{8(am_b)^3} + c_2 \frac{i}{8 (am_b)^3}(\nabla \cdot \tilde{E}  - \tilde{E} \cdot \nabla) \nonumber \\
&& -c_3 \frac{1}{8 (m_b)^2} \sigma \cdot (\nabla \times \tilde{E}  - \tilde{E} \times \nabla) - c_4 \frac{1}{2 am_b} \sigma \cdot \tilde{B} \nonumber \\
&& +  c_5 \frac{\Delta^{(4)}}{24 a m_b} - c_6 \frac{(\Delta^{(2)})^2}{16 (am_b)^2}, \\ \nonumber
\end{eqnarray}
with $c_{1..6}$ as the tuned improvement coefficients \cite{PhysRevD.85.054509}.
The bottom quark mass is tuned by setting the spin-averaged bottomonia kinetic mass ${1\over 4}(3M_{\Upsilon}+M_{\eta_b})$ to its experimental value.

The dibaryon states that we study in this work are constructed from the following single baryon operators:
\begin{eqnarray}\label{eq:singbar}
 	\Omega(llQ)_\alpha &=& \epsilon^{efg}\ \ l^e_\alpha (x) \ l^f_{\beta}(x)  (C\gamma_5)_{\beta \gamma} Q^g_{\gamma}(x), \\ \nonumber
 	\Omega(QlQ)_\alpha &=& \epsilon^{efg} \ \ Q^e_\alpha (x) \ l^f_{\beta}(x)  (C\gamma_5)_{\beta \gamma} Q^g_{\gamma}(x). \\ \nonumber 	
\end{eqnarray}
The Greek letters denote the spinor indices and the labels ($e,f,g$) are color indices.
The two flavor labels are denoted by ($l,Q$), where the label $l$  generally represents a lighter quark flavor and we vary that over a range, $l \in (s,c)$, and $Q$ is a heavier flavor such that $Q \in \{c,b\}$ and is   always heavier than $l$. 
The label $(s,c,b)$ denote strange, charm and bottom flavors respectively, and are set at their physical values.

From these operators, three types of dibaryon operators can be constructed with flavor symmetic $(FS)$ and spin zero  $(S=0)$ quantum numbers :
\begin{equation}\label{eq:dibops}
	(FS,S=0) = 
		\begin{cases}
		 \Omega(llQ)_{\alpha} \ (C \gamma_5)_{\alpha \beta} \ \Omega(llQ)_{\beta}  ,  \\
		\sqrt{2}\; \Omega(llQ)_{\alpha} \ (C \gamma_5)_{\alpha \beta} \ \Omega(QlQ)_{\beta} ,  \\
		\Omega(QlQ)_{\alpha} \ (C \gamma_5)_{\alpha \beta} \ \Omega(QlQ)_{\beta} . \\
		\end{cases}
\end{equation}
In the two-flavor isospin limit, these operators correspond to the nucleon-nucleon states in the spin-singlet channel and are degenerate.
Since, in this work, various two-flavor combinations of the dibaryon operators will be studied, and furthermore, since all of them are far away from the isospin limit,  there will be a large breaking of the degeneracy and as such these states can be considered almost independent.
The various flavor combinations of dibaryons that we investigate in this work, at the relevant physical  quark masses, are listed in Table~\ref{tab:diblist}. The possible two particle allowed energy levels are shown by the middle column while the lowest non-interacting (NI) states are shown by the last column. To be noted here that the lowest NI-energy levels need to be determined at the given quark masses of dibaryons as these NI-energy levels change their relative hierarchy with respect to quark masses. We elaborate this in the next section.

\begingroup
\renewcommand*{\arraystretch}{1}
\begin{table}
\begin{tabular}{c|c|c}\hline \hline
	 Dibaryon & Possible two baryon & Lowest \\ states & non-interacting (NI) states & NI states  \\ \hline \hline 
	  $\Omega_{bcb} \Omega_{bcb}$ & ($\Omega^{\frac{3}{2}}_{ccb}, \Omega^{\frac{3}{2}}_{bbb} $), $2\times \Omega^{\frac{3}{2}}_{bcb} $, $2 \times \Omega^{\frac{1}{2}}_{bcb} $ & $\Omega^{\frac{3}{2}}_{ccb}, \Omega^{\frac{3}{2}}_{bbb} $ \\ \hline 
	  
	  $\Omega_{ccb} \Omega_{bcb}$ & ($\Omega^{\frac{3}{2}}_{ccc},\Omega^{\frac{3}{2}}_{bbb} $), ($\Omega^{\frac{1}{2}}_{ccb},\Omega^{\frac{1}{2}}_{bcb} $),($\Omega^{\frac{3}{2}}_{ccb},\Omega^{\frac{3}{2}}_{bcb} $) & $\Omega^{\frac{3}{2}}_{ccc}, \Omega^{\frac{3}{2}}_{bbb} $ \\ \hline 
	 
	  $\Omega_{ccb} \Omega_{ccb}$ & ($\Omega^{\frac{3}{2}}_{ccc}, \Omega^{\frac{3}{2}}_{cbb} $), $2\times \Omega^{\frac{3}{2}}_{ccb} $, $2 \times \Omega^{\frac{1}{2}}_{ccb} $ & $\Omega^{\frac{3}{2}}_{ccc},\Omega^{\frac{3}{2}}_{cbb} $ \\ \hline  
	  
	  $\Omega_{bb} \Omega_{bb}$ & ($\Omega^{\frac{3}{2}}_{b}, \Omega^{\frac{3}{2}}_{bbb} $),  ($\Omega^{\frac{1}{2}}_{bb}, \Omega^{\frac{1}{2}}_{bb} $), ($\Omega^{\frac{3}{2}}_{bb}, \Omega^{\frac{3}{2}}_{bb} $) & $\Omega^{\frac{3}{2}}_{b}, \Omega^{\frac{3}{2}}_{bbb} $ \\ \hline
	  
	  $\Omega_{b} \Omega_{bb}$  & ($\Omega^{\frac{3}{2}}_{sss}, \Omega^{\frac{3}{2}}_{bbb}$), ($\Omega^{\frac{1}{2}}_{b}, \Omega^{\frac{1}{2}}_{bb} $), ($\Omega^{\frac{3}{2}}_{b}, \Omega^{\frac{3}{2}}_{bb} $) & $\Omega^{\frac{3}{2}}_{sss}, \Omega^{\frac{3}{2}}_{bbb}$  \\ \hline 	
	  
	  $\Omega_{cc} \Omega_{cc}$ & ($\Omega^{\frac{1}{2}}_{cc}, \Omega^{\frac{1}{2}}_{cc}$),($\Omega^{\frac{3}{2}}_{c}, \Omega^{\frac{3}{2}}_{ccc} $),($\Omega^{\frac{3}{2}}_{cc}, \Omega^{\frac{3}{2}}_{cc}$)  & $\Omega^{\frac{1}{2}}_{cc}, \Omega^{\frac{1}{2}}_{cc}$ \\ \hline 
	  
	  $\Omega_{c} \Omega_{cc}$  & ($\Omega^{\frac{1}{2}}_{c}, \Omega^{\frac{1}{2}}_{cc}$), ($\Omega^{\frac{3}{2}}_{c}, \Omega^{\frac{3}{2}}_{cc}$), ($\Omega^{\frac{3}{2}}_{sss}, \Omega^{\frac{3}{2}}_{ccc}$ )& $\Omega^{\frac{1}{2}}_{c}, \Omega^{\frac{1}{2}}_{cc}$ \\  \hline \hline											
\end{tabular}
\caption{\label{tab:diblist}List of physical dibaryon states (first column) studied in this work. 
The middle column shows the possible dibaryons with the same quark contents while the last column shows the lowest non-interacting levels. See the results section for details.}
\end{table}
\endgroup

\section{Results\label{sec:results}}
\begin{figure*}[ht]
{\hspace*{-0.3in}}
\includegraphics[width=0.51\textwidth,height=0.33\textwidth]{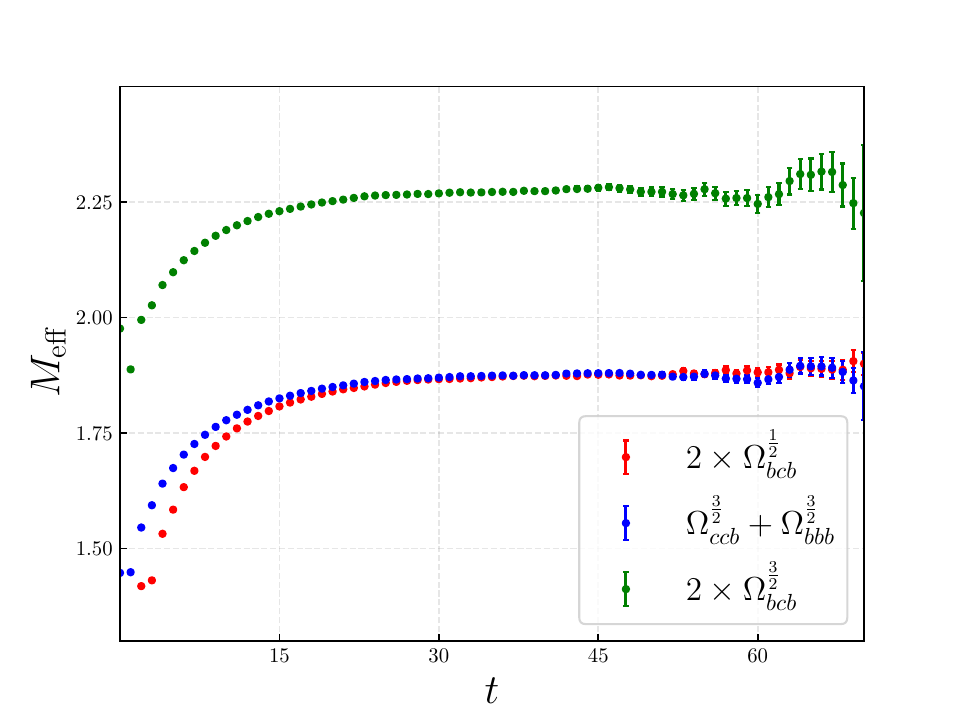}
\includegraphics[width=0.51\textwidth,height=0.33\textwidth]{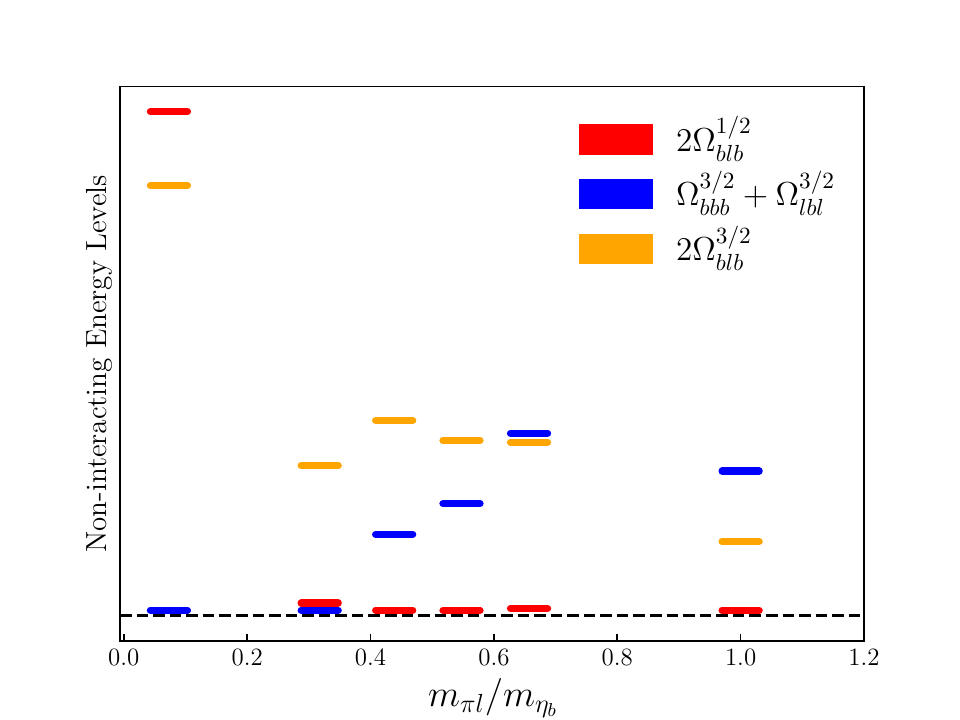}
\caption{\label{fig:NIlevel} Left panel : Comparison of non-interacting (NI) energy levels of two $\Omega^{\frac{1}{2}}_{bcb}$ baryons and combination of $\Omega^{	\frac{3}{2}}_{bbb}$ and $\Omega^{\frac{3}{2}}_{ccb}$ baryons. Data in red represents the effective mass of the product of two $\Omega^{\frac{1}{2}}_{bcb}$ correlators while the data in blue shows the effective mass of the product of $\Omega^{	\frac{3}{2}}_{bbb}$ and $\Omega^{\frac{3}{2}}_{ccb}$ correlators. See text for details. Right panel : Relative energy hierarchy of NI-energy levels for $\Omega_{bcb} \Omega_{bcb}$ with increase of quark mass $m_l$ shown by the ratio $m_{\pi l}/m_{\eta_b}$, where $m_{\pi l}$ is the pseduoscalar meson mass of the quark $l$, and $m_{\eta_b}$ is the mass of $\eta_b$.}
\end{figure*}
In this section, the analysis of the NI-energy levels, shown in Table~\ref{tab:diblist}, and the lowest energy states of dibaryons, shown in Eq. (\ref{eq:dibops}), are presented.  The NI-energy levels of dibaryons with heavy quarks  exhibit a different pattern in comparison to their light counterparts and must be fully understood before comparing their ground states to their lowest NI-energy levels.
This was already evident in the works in Ref. \cite{Junnarkar:2019equ, Junnarkar:2022yak} and are relevant in this work for identifying the lowest NI-energy level for a two-baryon systems with allowed spin parity.  
The analysis is therefore started by looking the relevant NI-energy levels as discussed below.

The interpolating operators that we 
utilized to compute the two-point correlation functions are as in
Eq.~(\ref{eq:singbar}). 
The two-point function for a given operator, ${\cal{O}}$, with a source (at time $t_{i}$) sink (at time $t_{f}$) setup is computed as:
\begin{align}\label{eq:corr_fun}
C_{\mathcal{O}}(t_i,t_f) = \sum_{\vec{x}} e^{-i\vec{p}.\vec{x}}\langle 0 | \mathcal{O}(\vec{x},t_f) \sum_{x_i}{\bar{\mathcal{O}}(\vec{x}_i,t_i)}|0 \rangle.
\end{align}
The spatial sum over the source operator is the consequence of employing zero-momentum projected wall source on the propagators.
It is convenient to compute effective masses from the correlators as, $M_{\text{eff}}(\tau) = \text{ln} \big(C(\tau)/C(\tau+1)\big)$,  to identify a plateau corresponding to the ground state in the limit of large temporal source-sink separation, (large $\tau = t_f - t_i$).
\begin{figure}[ht]
\includegraphics[width=0.45\textwidth]{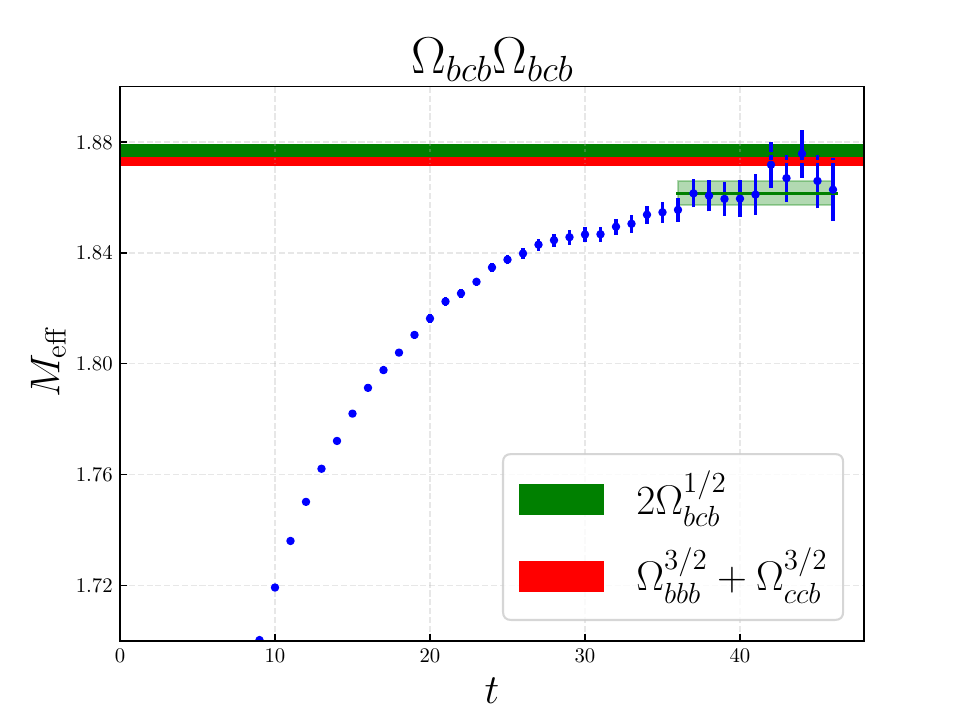}
\includegraphics[width=0.45\textwidth]{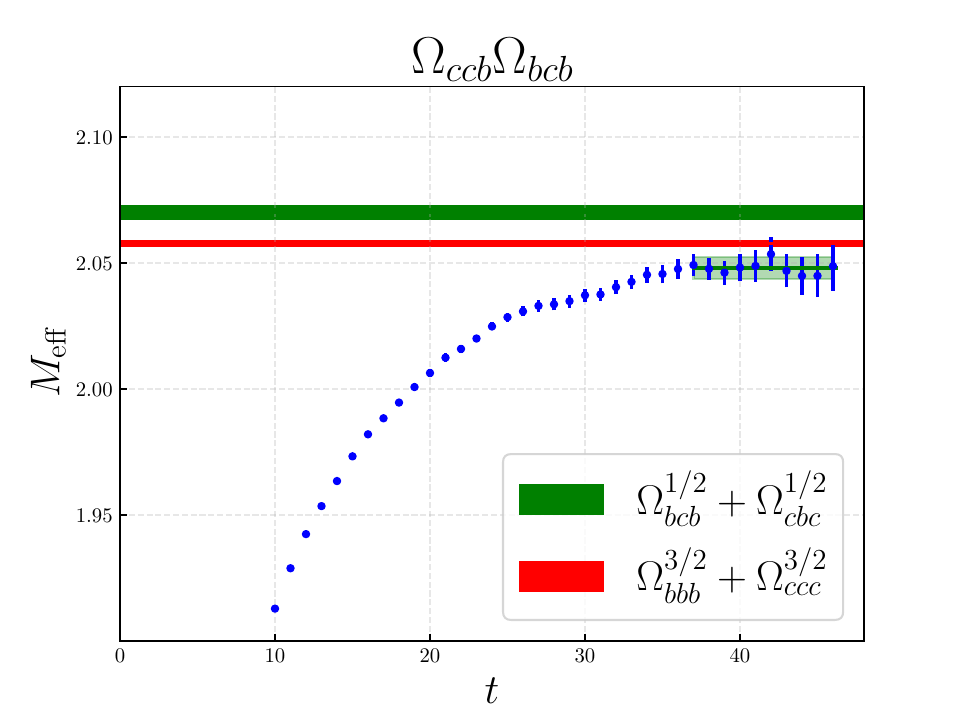}
\includegraphics[width=0.45\textwidth]{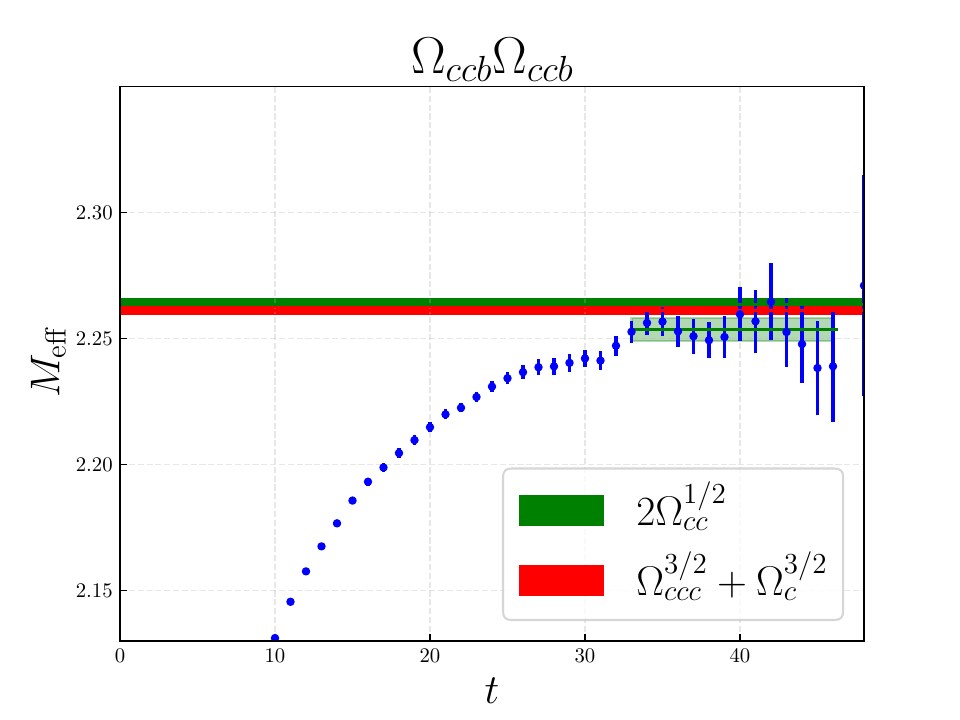}
\caption{\label{fig:omcb} Effective mass plots of $\Omega_{bcb} \Omega_{bcb}$, $\Omega_{ccb} \Omega_{bcb}$ and $\Omega_{ccb} \Omega_{ccb}$ dibaryons at lattice spacing $a=0.0582$ fm. The energy bands in green and red are the lowest noninteracting energy levels  of the spin-1/2 and spin-3/2 baryons respectively.}
\end{figure}
Similar to the case of heavy deuteron in Ref.~\cite{Junnarkar:2019equ}, the relevant spin-3/2 NI-energy levels composed of two Omega baryons are also considered which in some cases are found to be the lowest non-interacting energy levels.
In all cases, there is only one level composed of two spin-1/2 Omega baryons and two levels either of two  spin-3/2 Omega baryons or of two two-flavor spin-3/2 baryons, as are shown in Table~\ref{tab:diblist}.
In each dibaryon channel, all three levels are computed, and the lowest NI-energy level is identified accordingly and is shown in the third column of Table~\ref{tab:diblist}.
To find possible existence of a bound state we compare such lowest NI-energy levels
with the ground state of dibaryons obtained with the operators mentioned in Eq. (\ref{eq:dibops}).
We look at these NI-energy levels in more detail for the case of bottom-charm dibaryons which are listed in the first three rows of Table \ref{tab:diblist}.
\begin{figure*}[ht]
\includegraphics[width=0.45\textwidth]{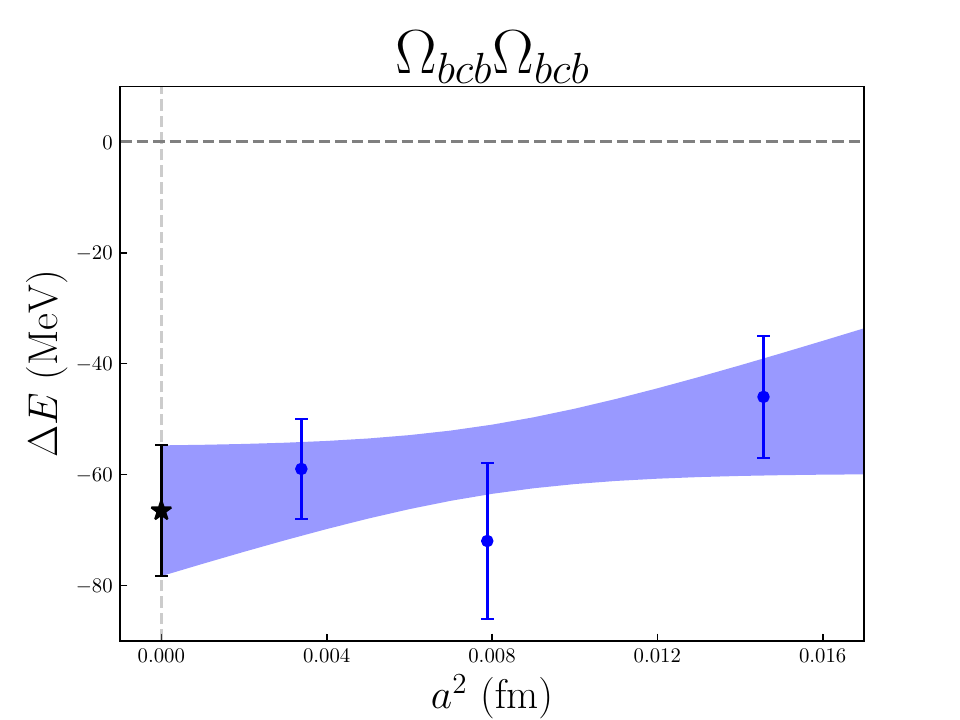}
\includegraphics[width=0.45\textwidth]{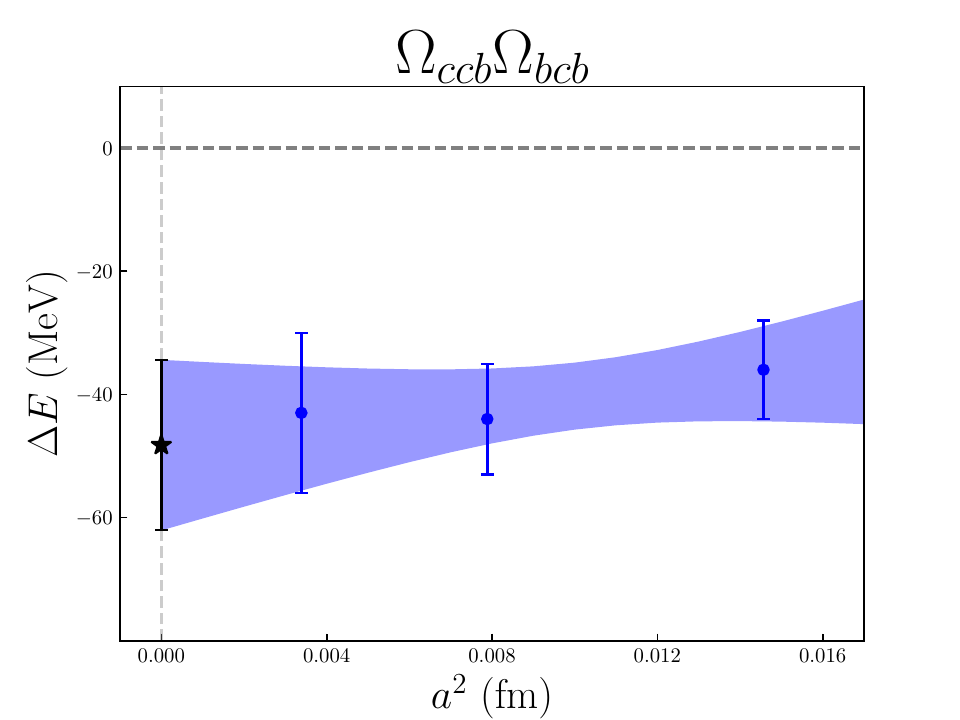}
\includegraphics[width=0.45\textwidth]{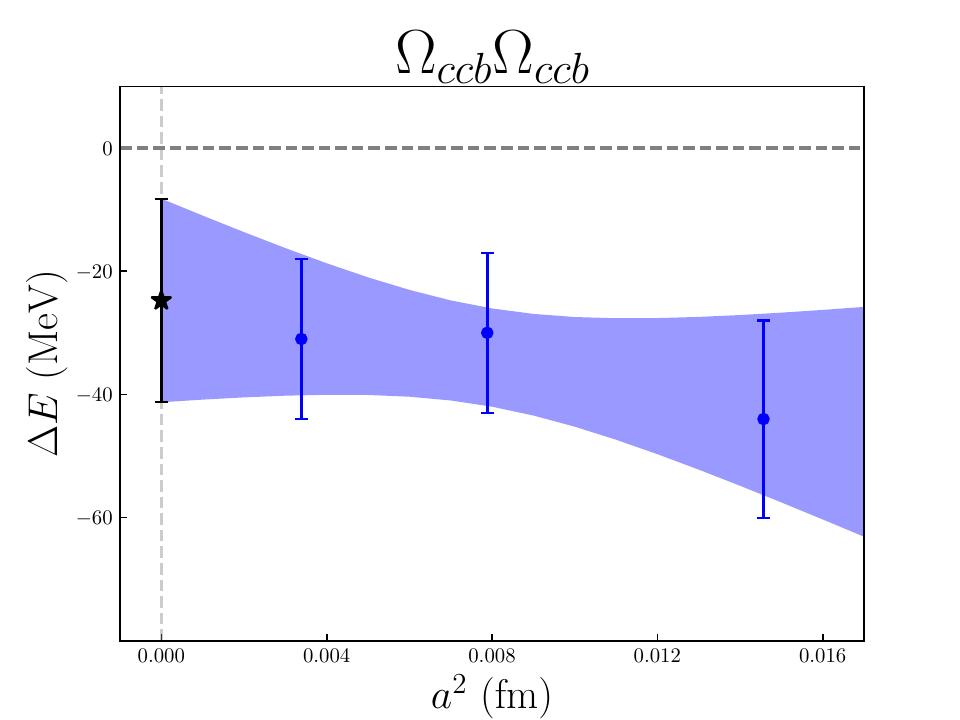}
\includegraphics[width=0.45\textwidth]{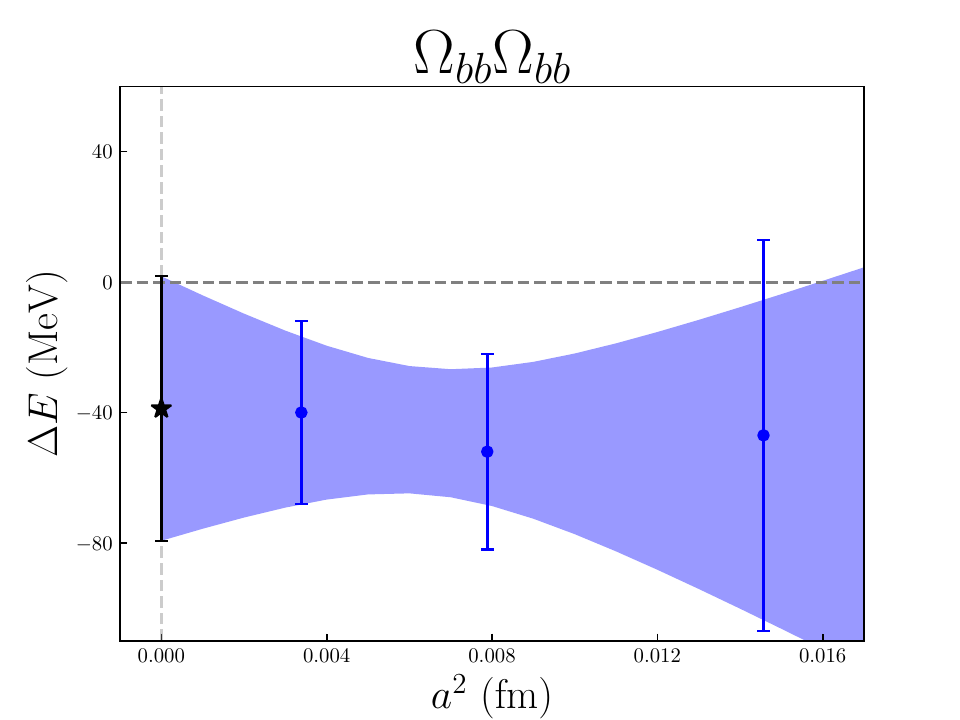}
\caption{\label{fig:cont_extrp_bottom} Results of continuum extrapolation of the extracted energy difference, $\Delta E$ (Eq. \ref{eq:DeltaE}), of various dibaryons in the bottom-charm and bottom-strange sectors.}
\end{figure*}

For the dibaryon state $\Omega_{bcb} \Omega_{bcb}$, the relevant NI single baryons are  $\Omega^{\frac{1}{2}}_{bcb}$, $\Omega^{\frac{3}{2}}_{ccb}$ and $\Omega^{\frac{3}{2}}_{bbb}$. 
Since there is no experimental determination of these states, we use  recent lattice QCD results from Refs. \cite{Brown:2014ena,Mathur:2018epb} to obtain their ground state masses. As indicated in the first row of Table \ref{tab:diblist}, their masses are:
	\begin{eqnarray}
2 \times \Omega^{\frac{1}{2}}_{bcb} &=& 22388(10) \ \text{MeV} , \\ \nonumber
\Omega^{\frac{3}{2}}_{bbb} + \Omega^{\frac{3}{2}}_{ccb} &=& 22392(11) \ \text{MeV} , \\ \nonumber 
2 \times \Omega^{\frac{3}{2}}_{bcb} &=& 22422(12) \ \text{MeV} .
	\end{eqnarray}
	Note that the non-interacting states of $2 \times \Omega^{\frac{1}{2}}_{bcb}$ and $\Omega^{\frac{3}{2}}_{bbb} + \Omega^{\frac{3}{2}}_{ccb}$ are close enough and will be difficult to distinguish them from their masses. We confirm this in our calculation by presenting effective mass plots of the states as shown in the left panel of Fig~\ref{fig:NIlevel}. 
	The plot shows effective mass of product of correlation function constructed as:
	\begin{align}
		C(\tau) = C_{B1}(\tau) \times C_{B2}(\tau) ,
	\end{align}
	where $B1$ and $B2$ are the relevant single baryons.
	The result in red is the effective mass of the product correlator of two $\Omega^{\frac{1}{2}}_{bcb}$ baryons and the result in blue is that of effective mass of the product correlator of $\Omega^{\frac{3}{2}}_{bcb}$ and $\Omega^{\frac{3}{2}}_{bbb}$ baryon correlators.
	The results are shown at the lattice of $a = 0.0582$ fm. 
	 The third NI-energy level for the two spin-3/2 baryons $\Omega^{\frac{3}{2}}_{bcb}$, shown with data in green, lies above the other two NI-energy levels.

For the dibaryon $\Omega_{ccb} \Omega_{bcb}$, the NI-energy levels for this flavor combination are shown in the second row of the Table \ref{tab:diblist}.  
 In this case, the lowest NI-energy level is clearly identified as the ($\Omega^{\frac{3}{2}}_{bbb},\Omega^{\frac{3}{2}}_{ccc}$) combination although the levels for spin-1/2 baryons are close by. 
The levels of the ($\Omega^{\frac{3}{2}}_{bcb},\Omega^{\frac{3}{2}}_{ccb}$ ) combination is identified to be the most well separated, and is found to be the highest one and hence is not relevant here.
It may be noted that the flavor structure of this dibaryon is the same in comparison to the spin one deuteron-like state $\mathcal{D}_{bc}$ studied in Ref.~\cite{Junnarkar:2019equ}.
The NI-energy levels are also the same and have been discussed in ~\cite{Junnarkar:2019equ}	 as well. 

For the dibaryon $\Omega_{ccb} \Omega_{ccb}$, the relevant NI-energy levels are shown in the third row of Table \ref{tab:diblist} and their ground state masses are obtained from Refs \cite{Brown:2014ena,Mathur:2018epb}:
	\begin{eqnarray}
		2 \times \Omega^{\frac{1}{2}}_{ccb} &=& 16010(12) \ \text{MeV} ,\\ \nonumber
		\Omega^{\frac{3}{2}}_{ccc}+ \Omega^{\frac{3}{2}}_{cbb} &=&  16007(10) \ \text{MeV} ,\\ \nonumber
		2 \times \Omega^{\frac{3}{2}}_{ccb} &=& 16052(10) ,\ \text{MeV}
	\end{eqnarray}
Similar to the case of $\Omega_{bcb} \Omega_{bcb}$, we find overlapping levels here as well and since the level from the state $\Omega^{\frac{3}{2}}_{ccc}+ \Omega^{\frac{3}{2}}_{cbb}$ is slightly lower than that of $2 \times \Omega^{\frac{1}{2}}_{ccb} $, we use the former as the lowest NI-energy level for comparing with the ground state masses of the dibaryon concerned. This hierarchy of NI-energy levels result is also confirmed by our data.

In addition to considering dibaryons at physical charm and bottom quark masses, we also compute the ground state spectrum at several unphysically heavy quark masses, $m_c < m_l \le m_b$.
This analysis is crucial for exploring the dependence of possible binding energies on the quark masses and for identifying whether a bound state emerges at specific quark mass values.
As such, we present a summary of the relevant levels for the dibaryon $\Omega_{bcb} \Omega_{bcb}$ is shown in the right panel of Fig~\ref{fig:NIlevel}.
 On the $x$-axis, a ratio of pseudoscalar meson mass ($m_{\pi l}$) to the $\eta_b$ meson mass ($m_{\eta_b}$) is presented. 
 We vary the pseudoscalar meson mass over a wide range, starting from kaon (corresponding to strange flavor) to  $\eta_b$ meson mass (representing bottom). 
For $\Omega_{bcb} \Omega_{bcb}$,  the NI-energy levels exhibit a pattern of level-shifting around the charm quark mass. 
Below the charm quark, the lowest NI-energy level is the two spin-3/2 baryons while above the charm quark mass, the lowest level is two spin-1/2 baryons. 
At the charm quark mass, the two levels, i.e, spin-1/2 and spin-3/2, are  close enough to not be distinguishable within the uncertainties of the lattice data.
In the charm sector, we take the lowest levels to be comprised of two spin-1/2 baryons. 
Performing this analysis is essential for correctly identifying the lowest NI-energy level, which is critical for determining the presence of a possible bound state in the heavy dibaryon spectrum.

\begin{figure*}[ht]
\includegraphics[width=0.45\textwidth]{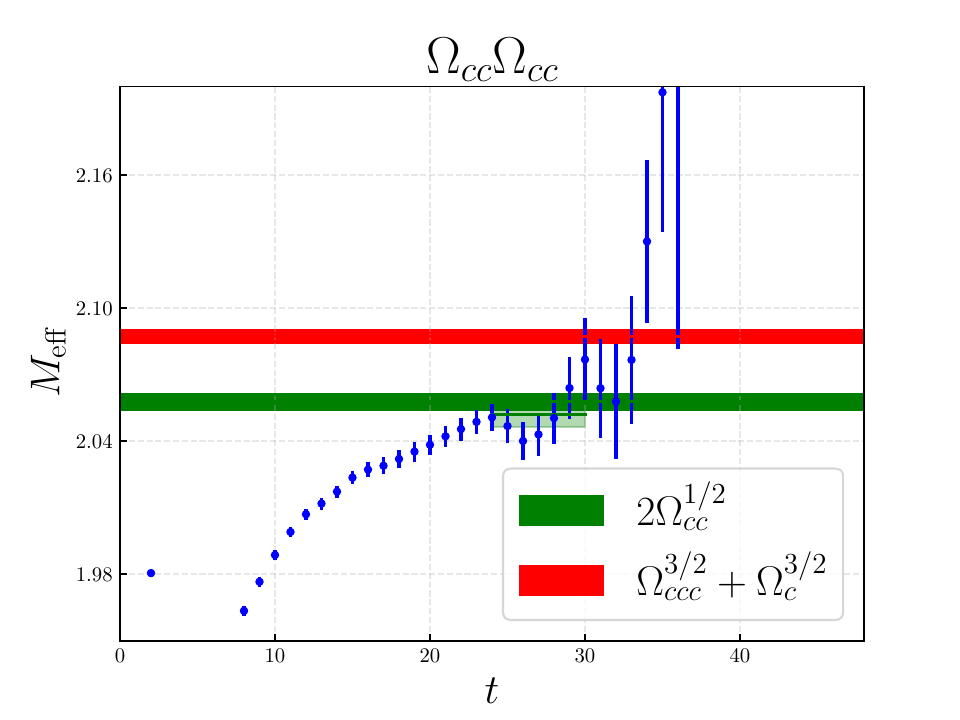}
\includegraphics[width=0.45\textwidth]{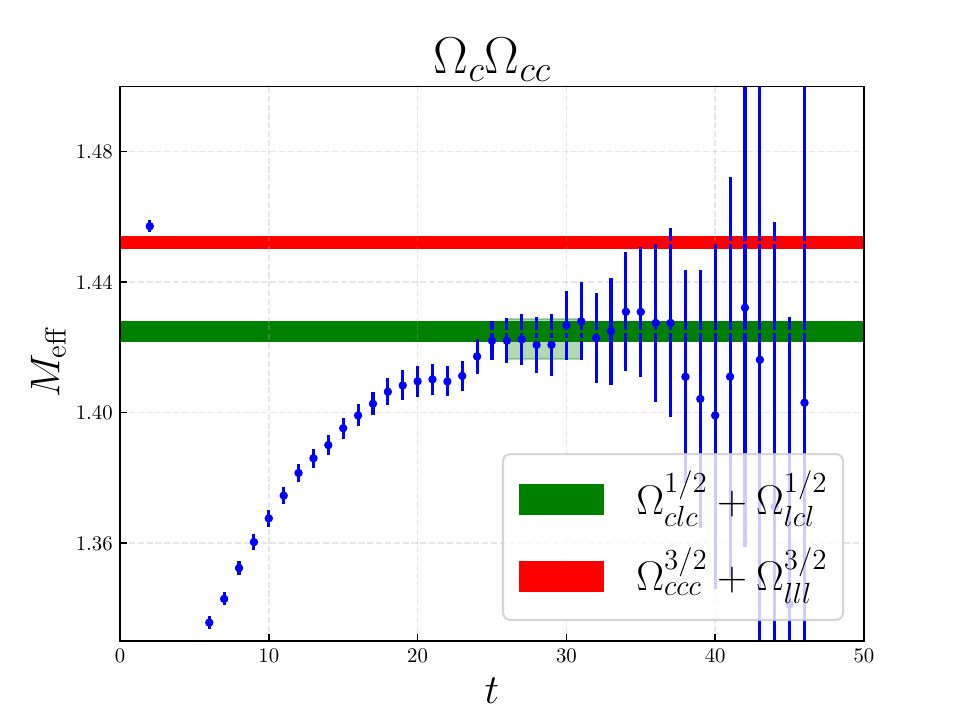}
\includegraphics[width=0.45\textwidth]{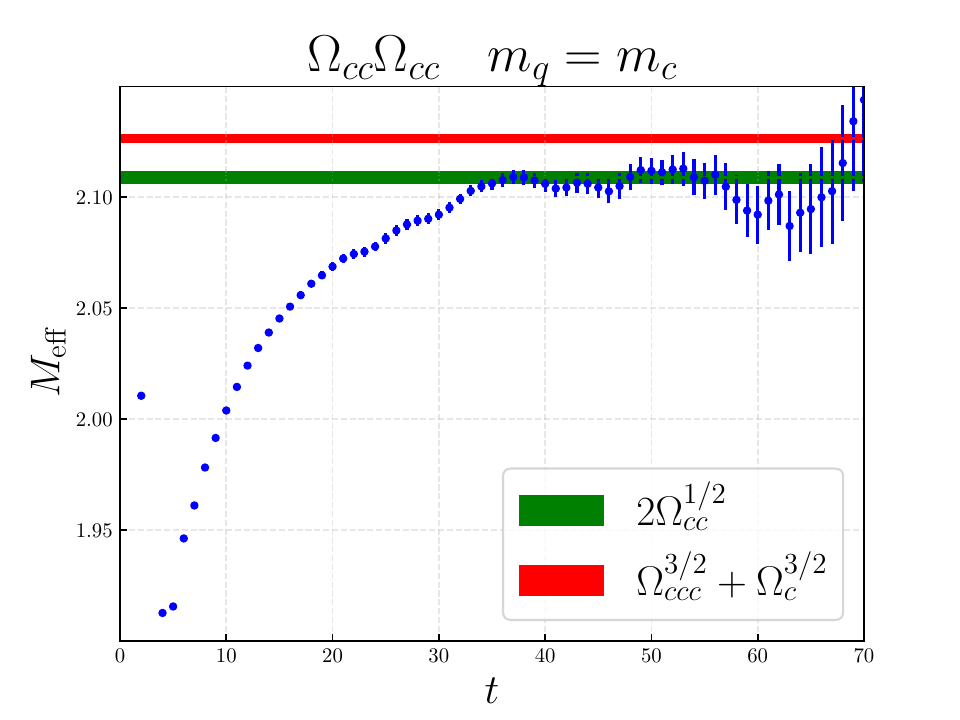}
\caption{\label{fig:omcc} Effective mass plots for dibaryons $\Omega_{cc} \Omega_{cc}$ (top left) and $\Omega_{c} \Omega_{cc}$ (top right) at lattice spacing $a=0.0583$ fm. 
The energy bands in green and red are the lowest noninteracting energy levels  of the spin-1/2 and spin-3/2 baryons, respectively. 
Bottom plot shows effective mass of $\Omega_{cc}\Omega_{cc}$ with $m_s = m_c$, at lattice spacing $a=0.0448$ fm (See text for explanation).}
\end{figure*}

\subsection{Ground state results}
\begingroup
\renewcommand*{\arraystretch}{1.5}
\begin{table}
\begin{tabular}{c|c|c|c|c} \hline \hline
	 \multirow{2}{*}{Dibaryon} & \multicolumn{3}{c|}{$\Delta E^a_{\Omega \Omega}$ (MeV)} & $\Delta E^{a=0}_{\Omega \Omega}$ \\ \cline{2-4} 
	&$a$ = 0.1206 & 0.0888 &0.0538 & (MeV)\\ \hline \hline
	  $\Omega_{bcb} \Omega_{bcb}$ & -46(11)& -72(14)& -59(9)& -66(11)  \\
	 $\Omega_{ccb} \Omega_{bcb}$ &-36(8) & -44(9)& -43(13) & -48(13) \\
	 $\Omega_{ccb} \Omega_{ccb}$ &-44(16) & -30(13)& -31(13)& -24(16)  \\ \hline \hline 
	$\Omega_{bb} \Omega_{bb}$ &-47(60) &-52(30) &-40(28) & -38(40)  \\
	 $\Omega_{b} \Omega_{bb}$ &-20(57) &-15(38) & -40(31)& -43(43)  \\ \hline \hline
\end{tabular}
\caption{\label{tab:ce_bottom} Continuum extrapolation of the dibaryon effective binding energies ($\Delta E$ (Eq.\ref{eq:DeltaE})) in the bottom flavor sector with three lattice spacings (in units of fm).}
\end{table}
\endgroup
In this subsection, the analysis of ground state correlation functions of the various dibaryons, shown in Table~\ref{tab:diblist}, is presented.  
The results of bottom-charm sector are analyzed first, followed by the bottom-strange sector.
For all the states considered here, the correlation functions are fitted to a single exponential of the form, $C(\tau) \sim A e^{-E_0 \tau}$, at large values of $\tau$.
This will be indicated on the effective mass plots of the correlators with a fit band which will also indicate the fit range.
From the fit, the energy difference, at each ensemble with lattice spacing $a$, are computed as:
\begin{equation}\label{eq:DeltaE}
\Delta E^a_{\Omega \Omega} = E^a_{\Omega \Omega} - E^a_{NI} .
\end{equation}
where $E^a_{\Omega \Omega}$ is the ground state energy obtained from the one exponential fit to the relevant dibaryon $(\Omega \Omega)$ two-point function (Eq. \ref{eq:dibops}). Similarly $E^a_{NI}$ is the  ground state energy of the corresponding lowest NI-energy level.
If this energy difference is found to be negative, then it will be termed as an  effective binding energy.
In addition to computing the energy differences as described, we also compute a ratio correlator of the dibaryons with respect to the relevant single baryons as,
\begin{equation}{\label{eq:ratiocorr}}
	\mathcal{R}(\tau) = \frac{C_{\Omega \Omega}(\tau)}{C_{B_1}(\tau) \times C_{B_2}(\tau)} \xrightarrow[\text{large} \ \tau]{} Ae^{-\Delta E^a_{\Omega \Omega} \tau},
\end{equation}
which provides an alternate way to  access the effective energy difference, $\Delta E^a_{\Omega \Omega}$, at large times. 
The analysis of such a ratio correlator however requires more care, since the leading exponential behavior of the correlators in the numerator and denominator can be different due to presence of different asymptotic states and the effective energy difference can produce a misleading plateau at short distance. Only at a very large times such a method would be reliable.
In the analysis here, the ratio correlator is only used as a cross-check to the direct method of computing the energy differences as in Eq. (\ref{eq:DeltaE}).
With these descriptions, the dibaryon analysis can now be proceeded in the next two subsections.
\begin{figure*}[ht]
\includegraphics[width=0.45\textwidth]{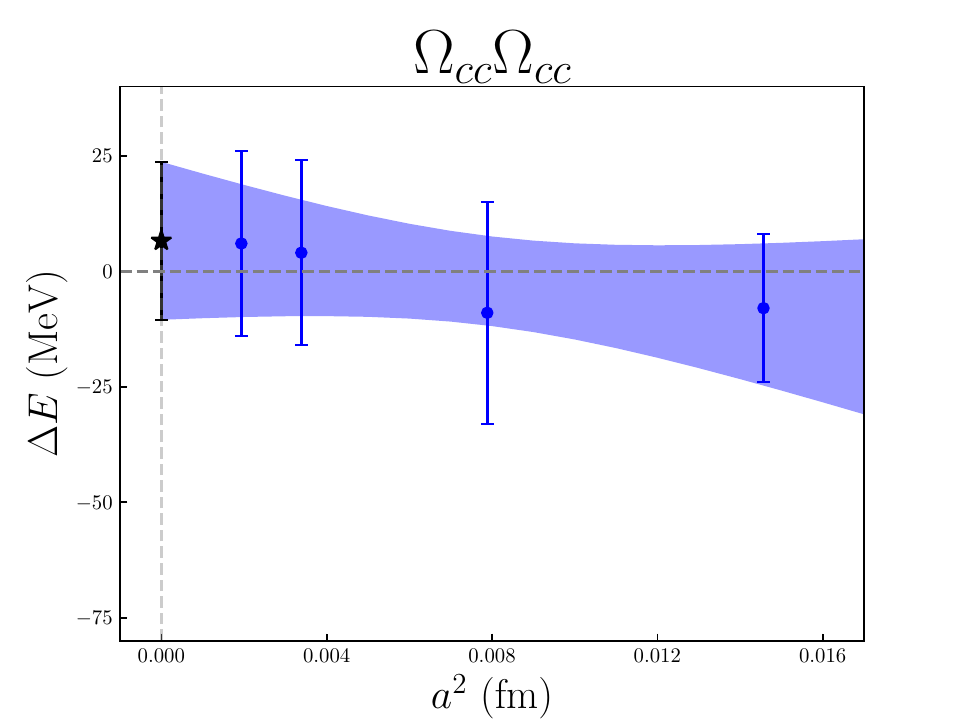}
\includegraphics[width=0.45\textwidth]{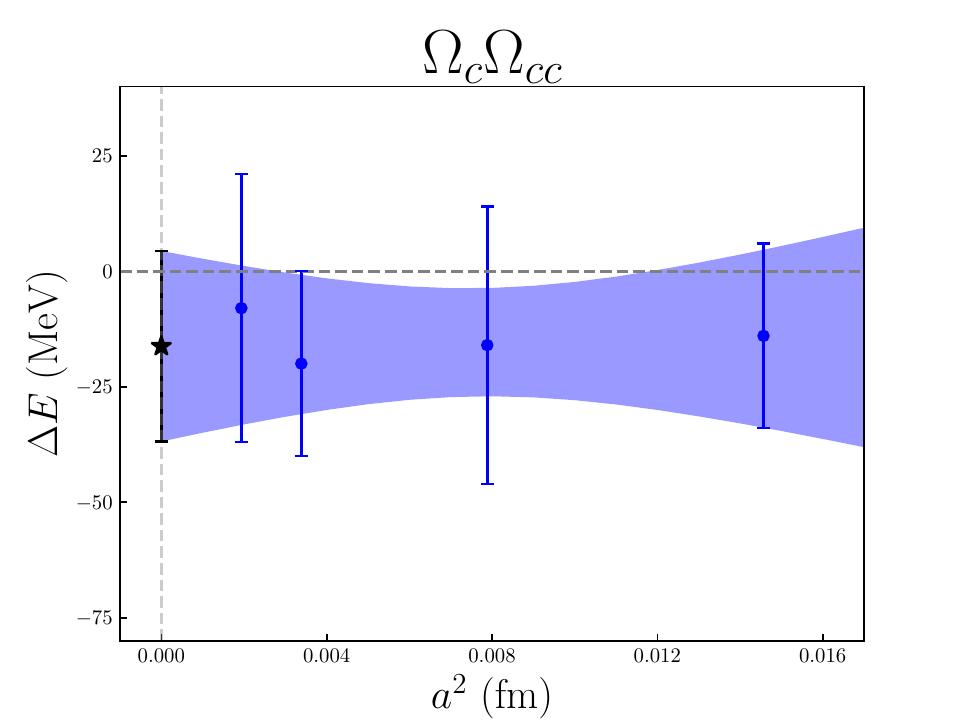}
\caption{\label{fig:cont_extrp_charm} Results of continuum extrapolation of the extracted energy difference, $\Delta E$ (Eq. \ref{eq:DeltaE}), of dibaryons in the charm sector.}
\end{figure*}
\begingroup
\renewcommand*{\arraystretch}{1.5}
\begin{table}
\begin{tabular}{c|c|c|c|c|c} \hline \hline
	 \multirow{2}{*}{Dibaryon} & \multicolumn{4}{c|}{$\Delta E^a_{\Omega \Omega}$ (MeV)} & $\Delta E^{a=0}_{\Omega \Omega}$  \\ \cline{2-5} 
	& $a=0.1206$ & 0.0888 &0.0538 & 0.0448 &(MeV)\\ \hline\hline 
	$\Omega_{cc} \Omega_{cc}$ &-8(16) &-9(24) &4(22) & 6(20)& 6(17)  \\
	$\Omega_{c} \Omega_{cc}$ &-14(20) &-16(30) & -20(20)& -8(29) & -16(20) \\
	\hline \hline
	\end{tabular}
\caption{\label{tab:ce_charm} Continuum extrapolation of the dibaryon effective binding energies ($\Delta E$ (Eq. \ref{eq:DeltaE})) in the charm sector with four lattice spacings (in units of fm).}
\end{table}
\endgroup
\begin{figure*}[ht]
\includegraphics[width=0.45\textwidth]{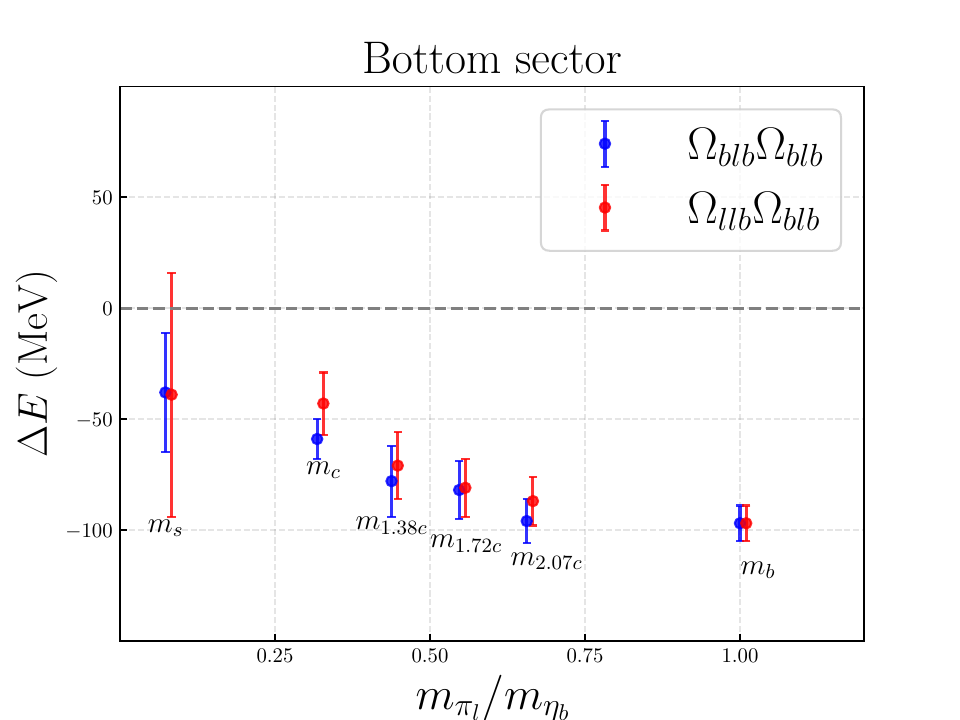}
\includegraphics[width=0.45\textwidth]{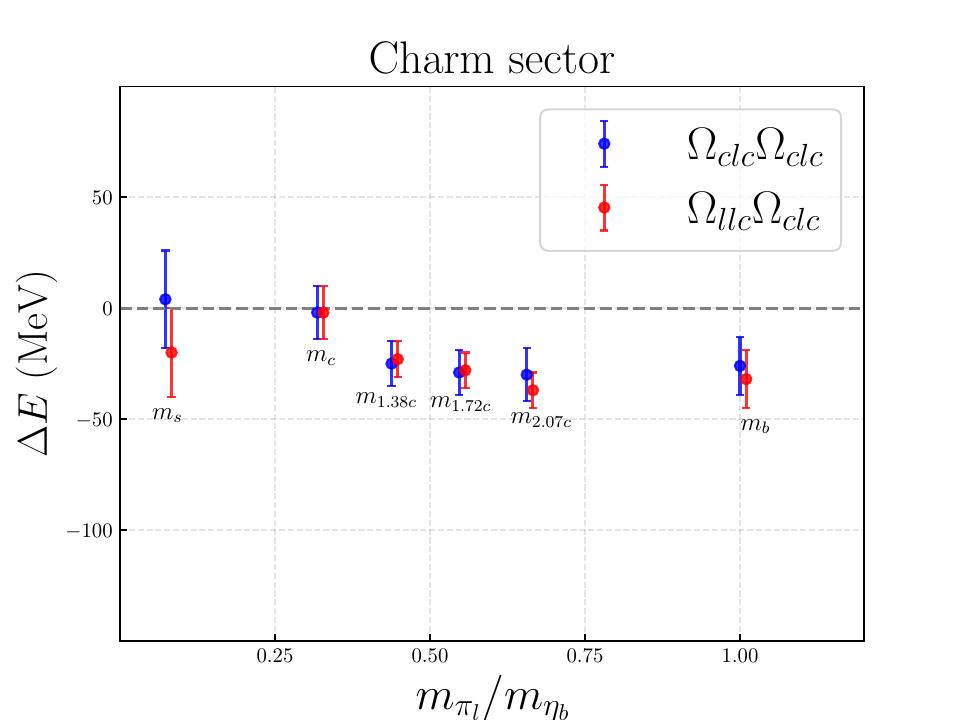}
\caption{\label{fig:heavy_mq} Heavy quark mass dependence of dibaryons $\Omega_{blb} \Omega_{blb}$ \& $\Omega_{llb} \Omega_{blb}$ in the bottom sector (left panel) and $\Omega_{clc} \Omega_{clc}$ \& $\Omega_{llc} \Omega_{clc}$ in the charm sector (right panel). The flavors bottom ($b$) and charm ($c$) are fixed to their respective physical values and the flavor $l$ varies from strange to bottom quark masses represented by the ratio $m_{\pi l}/m_{\eta_b}$, where $m_{\pi l}$ is the pseudoscalar meson mass of the quark flavor $l$.}
\end{figure*}
\subsubsection{Results in the bottom sector}
\noindent{\bf Bottom-charmed:} There are three physical states here that are given in the first row of Table~\ref{tab:diblist}.

\noindent{\bf A.} $\Omega_{bcb} \Omega_{bcb}$: 
The effective mass plot of the ground state of $\Omega_{bcb} \Omega_{bcb}$ dibaryon, for the ensemble with lattice spacing $a = 0.0582$ fm, is shown in the top left panel of the Fig.~\ref{fig:omcb}.
As noted earlier, the NI-energy levels are expected to overlap which are shown in the red and green fit bands. 
Since, in this case, the spin-3/2 level seems just slightly lower than the two spin-1/2 levels, we consider it as the lowest NI-energy level and the dibaryon correlator is compared to it.

The dibaryon correlation function is fitted at large source-sink separation from $\tau \sim 2.2$ fm  to $\tau \sim 2.8$ fm, which is a large enough separation for the ground state to be dominant, and that is indicated with a light-green band. 
Since dibaryon level lies below the NI-energy level, this is an indication of the presence of a  bound state. The extracted energy difference of the ground state from the lowest NI-energy level, $\Delta E^a$, is shown in the first row of Table~\ref{tab:ce_bottom}.
Similar results on $\Delta E^a$ are also extracted at two other lattice spacings and are tabulated in Table~\ref{tab:ce_bottom}.
A continuum extrapolation of these results is then performed with a linear ansatz in the square of lattice spacing as:
\begin{equation}\label{eq:cont_extrp}
	\Delta E^a_{\Omega \Omega} = \Delta E^{a=0}_{\Omega \Omega} \ + c_2 \, a^2. 
\end{equation}
We have taken such a form as the overlap action does not have ${\cal{O}}(ma)$ error and the leading error is expected to be ${\cal{O}}(ma)^2$, $ma$ being the lattice bare quark mass.
The continuum extrapolated result, $\Delta E^{a=0}$,  is shown in the top left panel of Fig.~\ref{fig:cont_extrp_bottom} and also in Table.~\ref{tab:ce_bottom}.

Note that no substantial lattice spacing dependence is observed since the slope is consistent with zero within $1\sigma$ errorbar.
This result indicates the possibility of a bound state with binding energy of
\begin{equation}
	\Delta E^{a=0}_{\Omega_{bcb} \Omega_{bcb}} = -66(11) \ \mathrm{MeV}.
\end{equation}

\noindent{\bf B.} $\Omega_{ccb} \Omega_{bcb}$: The effective mass corresponding to $\Omega_{ccb} \Omega_{bcb}$ dibaryon, is shown in the middle panel of Fig.~\ref{fig:omcb}. The ground state mass is clearly lie below the NI-energy level, which in this case is spin-3/2 $\Omega^{\frac{3}{2}}_{bbb} +\Omega^{\frac{3}{2}}_{ccc}$ baryons.
This again suggests the possibility of the existence of a bound state. 
The splitting between the two NI-energy levels is also found to be consistent with  the other lattice results~\cite{Brown:2014ena,Mathur:2018rwu}.
For other lattice spacings, we also find that the dibaryon levels lie below the lowest NI-energy levels.
After continuum extrapolation with the same fit form (Eq. \ref{eq:cont_extrp}), as shown in the top right panel of Fig.~\ref{fig:cont_extrp_bottom},
we find the possible binding energy to be
\begin{equation}
	\Delta E^{a=0}_{\Omega_{ccb} \Omega_{bcb}} = -48(13) \ \mathrm{MeV}.
\end{equation}
Here again, we find no dependence on lattice spacings.

\noindent{\bf C.} $\Omega_{ccb} \Omega_{ccb}$: The effective mass plot of the dibaryon $\Omega_{ccb} \Omega_{ccb}$ is presented in the bottom panel of Fig.~\ref{fig:omcb}, which again shows the existence of an energy level below the lowest NI-energy level. As in previous two cases, we calculate the 
$\Delta E_{\Omega_{ccb} \Omega_{ccb}}(a)$ at each lattice ensemble and then extrapolate the results to the continuum limit, as shown in the 
bottom left panel of Fig.~\ref{fig:cont_extrp_bottom}. In this case the continuum extrapolated value is found to be
\begin{equation}
	\Delta E^{a=0}_{\Omega_{ccb} \Omega_{ccb}} = -24(16) \ \mathrm{MeV}.
\end{equation}
All three cases of these dibaryons discussed above show the presence of  an energy level below their respective lowest NI-energy level.
Further, results in all three cases survive the continuum extrapolation making it a robust observation.
Together the results on the charmed-bottom dibaryons indicate the possibility of presence of bound states and suggest a trend where the effective binding energy ($\Delta E^{a=0}$) is found to be deeper for the dibaryon with the most number of heavier quarks. 
The case of  $\Omega_{bcb}\Omega_{bcb}$ being the most strongly bound.

\noindent{\bf Bottom-Strange:} The dibaryons with strange and bottom quarks, denoted as $\Omega_{bb} \Omega_{bb}$ and $\Omega_{b} \Omega_{bb}$, are shown in the 4th and 5th rows of Table~\ref{tab:diblist}. Here again we follow the same procedure to extract $\Delta E^{a}$ at each lattice ensemble which are then extrapolated to the continuum limit, as shown in the bottom right panel of Fig.~\ref{fig:cont_extrp_bottom}. 
In both the cases, although we find an energy level below their respective lowest NI-energy level, large uncertainties prohibit to make any conclusion about the existence of bound states. This is understandable since the strange quark being lighter such dibaryons require much higher statistics in order to provide any precise results. However, since these are interesting physical channels, further efforts with large statistics and perhaps with GEVP method~\cite{Luscher:1990ck} should be considered in future.

\begin{figure*}[ht]
\includegraphics[width=0.45\textwidth]{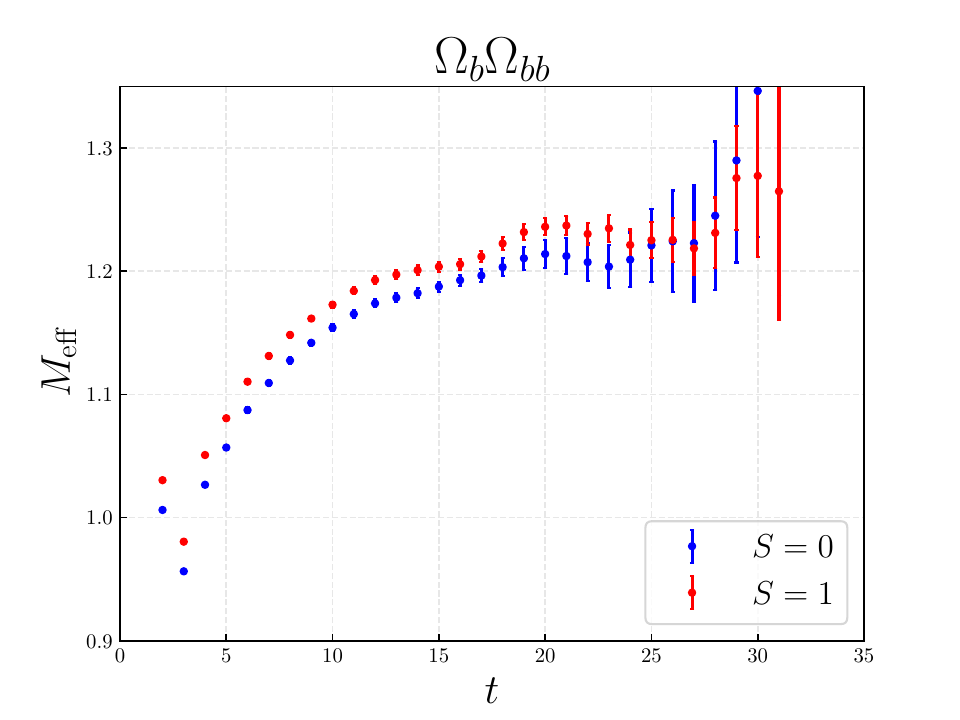}
\includegraphics[width=0.45\textwidth]{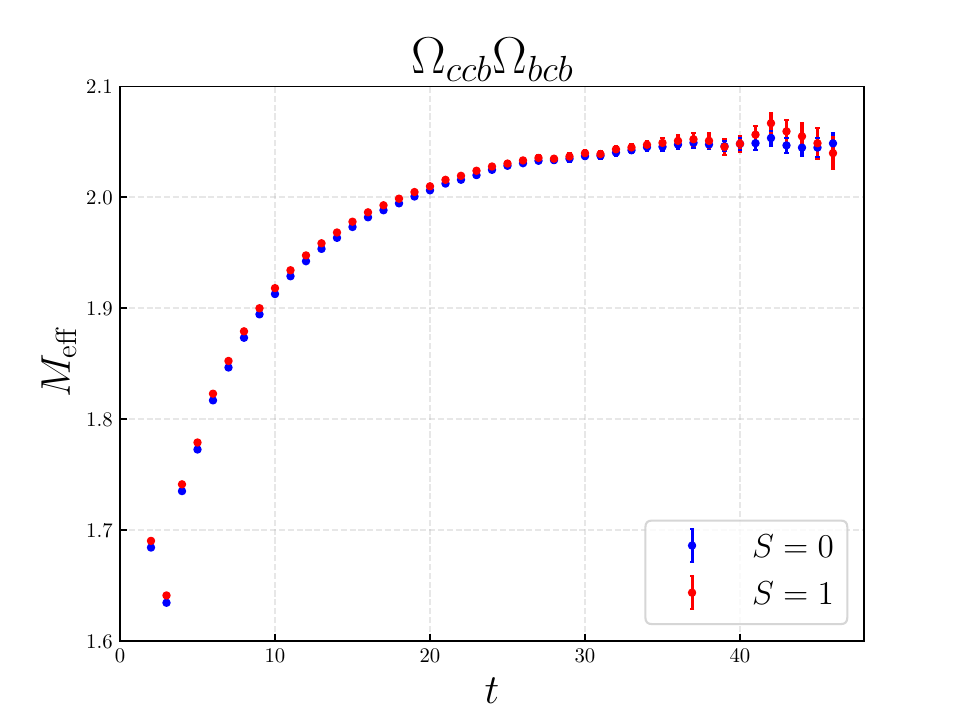}
\caption{\label{fig:hqs} Comparison of effective mass plots between spin-zero and spin-one states of heavy dibaryons. The spin-one states were computed in Ref.~\cite{Junnarkar:2019equ}. While there is a difference between the effective masses on the left panel showing breaking of heavy quark spin symmetry, the coincidence in the right panel suggests the exact heavy quark spin symmetry.}
\end{figure*}
\subsubsection{Results in the charm sector}
In the charm sector, the dibaryons (studied at physical quark masses) are listed in the bottom two rows of Table~\ref{tab:diblist}.
These are namely $\Omega_{cc} \Omega_{cc}$ and $\Omega_c \Omega_{cc}$ at physical charm and strange quark masses. 
The results of various states in this sector are shown in Fig.~\ref{fig:omcc}.

\noindent{$\Omega_{cc} \Omega_{cc}$:} The effective mass of $\Omega_{cc} \Omega_{cc}$ is shown in the upper left panel of Fig.~\ref{fig:omcc}.
In contrast with the $\Omega_{bb} \Omega_{bb}$ dibaryon, in the charm sector, the lowest NI-energy level is that of two spin-1/2 baryons, namely that of two $\Omega^{\frac{1}{2}}_{cc}$ baryons (see the  last two columns of Table ~\ref{tab:diblist})
The splitting between the spin-1/2 NI-energy level and spin-3/2 NI-energy level is found to be consistent with that of other lattice results~\cite{Brown:2014ena,Mathur:2018rwu}.
The effective mass of the $\Omega_{cc} \Omega_{cc}$ dibaryon is shown by blue-colored data which seems to plateau around the green-colored spin-1/2 NI-energy level. 
With the given statistics, it is unclear whether $\Omega_{cc} \Omega_{cc}$ represent a different energy level than that of spin-1/2 NI-energy level. 

In order to understand the trend in the energy levels for $\Omega_{cc} \Omega_{cc}$, we additionally calculate $\Omega_{cc} \Omega_{cc}$ with the light quark set at the charm mass i.e, $m_s = m_c$. If there is any bound state with such a quark configuration it may emerge at a higher quark mass, and hence we adopt such a choice.
To mitigate discretization effect we carry out this calculation with our finest lattice ensembles at  $a = 0.0448$ fm. The result is shown at the bottom  panel of Fig.~\ref{fig:omcc}. 
The effective mass in this case also seem to overlap with the lowest NI-energy level. This is possibly further suggesting there is  no energy level below the lowest NI-energy level for the $\Omega_{cc} \Omega_{cc}$ baryon at physical strange quark mass.
The trend  persists across the four lattice spacings and the result is shown in the  left panel of Fig.~\ref{fig:cont_extrp_charm} (and also in Table~\ref{tab:ce_charm}) with the final continuum extrapolated energy difference,
\begin{equation}
	\Delta E^{a=0}_{\Omega_{cc} \Omega_{cc}} = 6(17) \ \mathrm{MeV}.
\end{equation}

\noindent{$\Omega_{c} \Omega_{cc}$:} The plot in the upper right panel of Fig.~\ref{fig:omcc} shows the effective mass for the dibaryon $\Omega_{c}\Omega_{cc}$ which is also seen to be consistent with the lowest NI-energy level of two spin-1/2 baryons.
Within the statistical uncertainties the same trend of consistency between the $\Omega_{c}\Omega_{cc}$ and the lowest NI-energy level is also found at other three lattice ensembles.
The fitted energy differences are shown in the second row of Table~\ref{tab:ce_charm}. 
The continuum extrapolation with the fit form as in Eq. (\ref{eq:cont_extrp}) yields:
\begin{equation}
	\Delta E^{a=0}_{\Omega_{c} \Omega_{cc}} = -16(20) \ \mathrm{MeV},
\end{equation}
and also in the right panel of Fig.~\ref{fig:cont_extrp_charm}, suggesting no bound state with $\Omega_{c}\Omega_{cc}$ configuration.

\begingroup
\renewcommand*{\arraystretch}{1.2}
\begin{table}
\begin{tabular}{c|c|c|c|c|c}\hline \hline
\multirow{2}{*}{$m_l$} & \multirow{2}{*}{$m_{\pi l}/m_{\eta_b}$} & \multicolumn{4}{c}{$\Delta E^{a}$(MeV)} \\  \cline{3-6}  	
	&&$\Omega_{blb}\Omega_{blb} $ &  $\Omega_{llb}\Omega_{blb} $ & $\Omega_{clc}\Omega_{clc} $ &  $\Omega_{llc}\Omega_{clc}$ \\ \hline \hline 
        $m_s$       & 0.0732 & -38(40)& -39(55) & 4(22)   & -20(20) \\ \hline
        $m_c$       & 0.318 & -59(9) & -43(14) & -2(12)  & -2(12) \\ \hline
        $m_{1.38c}$ & 0.438 & -78(16) & -71(15) & -25(10) & -23(8) \\ \hline
        $m_{1.72c}$ & 0.547 & -82(13) & -81(13) & -29(10) & -28(8) \\ \hline
        $m_{2.07c}$ & 0.656 & -96(10) & -87(11) & -30(12) & -37(8) \\ \hline
        $m_{b}$     & 1.0 & -97(8)    & -97(8)  & -26(13) & -32(13) \\ \hline \hline
	
\end{tabular}
\caption{\label{tab:heavymq} Effective binding energies for the four dibaryons computed at heavier than physical charm quark masses at lattice spacing $a=0.0583$ fm.}
\end{table}

\subsection{Finite volume effects}
For these dibaryons we compute their ground state energy level, compare it with the respective lowest NI-energy level and calculate their difference, $\Delta E$.  
However, in order to associate $\Delta E$ with the binding energy one needs to carry our finite volume analysis of these energy levels. The interpretation of a bound state is understood to be a pole in the scattering amplitude which can be obtained by performing a Lüscher analysis~\cite{Luscher:1990ck}.
Here, there is a particular advantage of working  with heavy quark multi-hadron states.
This is associated with the suppression of  corrections to the infinite volume binding energy from the finite volume energy levels.
This can be understood by considering the finite volume corrections as:
\begin{eqnarray}\label{Eq:finite_vol_eff}
 \Delta_{FV} =  E_{FV} - E_{\infty} &\propto &  \mathcal{O}(e^{-k_{\infty} L })/L,\nonumber \\
   \mathrm{with}\quad k_{\infty} &=& \sqrt{(m_1 + m_2) B_{\infty}}\,.
\end{eqnarray}
where, $k_{\infty}$ is the binding momentum of the infinite volume state and $B_{\infty}$ is the binding energy of the state in infinite volume,  $E_{FV}$ is the finite volume energy level, and ($m_1, m_2$) are the masses of the two noninteracting hadrons with energy $m_1 + m_2$. 
As has been pointed out in Refs.~\cite{Junnarkar:2018twb,Junnarkar:2019equ,Junnarkar:2022yak},
in the case of multi-hadrons made of heavy quarks, the combination  $m_1+m_2$ of two heavy baryons is large enough to provide a large suppression implying that $\Delta_{FV}$ would expect to  be rather small. 
As a result, $\Delta E$ values will be expected to be closer  to the respective binding energies of dibaryons.
This will be particularly relevant for dibaryons in the bottom-charm sector, where there are clear indications of lowest energy levels far away from their respective lowest NI-energy levels, and the constituent baryons are quite heavy in masses.
However, as mentioned previously, to confirm these findings it is necessary to carry out a rigorous finite volume scattering amplitude analysis, particularly in the cases where the ground states are closer to the lowest NI-energy-levels. However, such an analysis is currently beyond the scope of this work.

\subsection{Heavy quark mass dependence}
In this subsection, the results of heavy quark mass dependence of the dibaryons in the bottom and charm sector are presented.
The variation of the extracted energy difference,  $\Delta E^{a=0}$, which can be interpreted as the binding energy if its survive a rigorous finite volume study, with respect to changes in quark masses is instructive in understanding the fate of dibaryons as bound states. To investigate the quark mass dependence of any possible binding energy, we compute the relevant dibaryon correlation functions, and compute $\Delta E$, over a wide range of quark masses, as shown in 
Table~\ref{tab:heavymq}. The dibaryons denoted in Table~\ref{tab:heavymq} have two flavor labels: heavy, which is either bottom ($b$) or charm ($c$),
and light, $l$, which is varied 
in between strange to bottom quark masses, $m_s \le m_l \le m_b$, as shown in the first column of Table~\ref{tab:heavymq}. 
The second column represents the ratio of the corresponding pseudoscalar meson mass ($m_{\pi l}$) relative to $\eta_b$ mass, which can be utilized as a dimensionless quantity for showing quark mass dependence. The results that we obtain for $\Delta E$ at those quark masses for various type of two-flavored spin-singlet dibaryon configurations are shown in the columns 3-6.

In particular, in the left panel of Fig.~\ref{fig:heavy_mq} we show  $\Delta E$ values for the dibaryons 
$\Omega_{blb} \Omega_{blb}$ and $\Omega_{llb}\Omega_{blb}$, at various quark masses represented by the ratio 
 $m_{\pi l}/m_{\eta_b}$.
For both the dibaryons, particularly for $\Omega_{llb}\Omega_{blb}$,  at the strange quark mass, the extracted $\Delta E$ values are consistent with zero within $1-2\sigma$ and hence no definitive conclusion can be drawn about the presence of a bound state. More statistics is needed for that.
However, at higher values of $m_l$, a clear trend of deepening in $\Delta E$ values is observed as the quark mass increases. Though a rigorous finite volume amplitude analysis is necessary to make a connection between the extracted  $\Delta E$ to the binding energy and associated pole structures, there is a strong indication of increase of binding energies considering the large values of $\Delta E$ and that they are far away from the non-interacting levels.
The trend of possible increment of binding energy with increase of quark masses is also similar to observed in Ref.~\cite{Junnarkar:2019equ}. 

Similar results on the heavy quark mass dependence of $\Delta E$ in the charm sector are shown in the right panel of Fig.~\ref{fig:heavy_mq}.
There are two dibaryons here: $\Omega_{clc} \Omega_{clc}$ and $\Omega_{llc}\Omega_{clc}$.
The trend in quark mass dependence is interesting in the following way :
At the strange quark mass, the results are, with the limited statistics, inconclusive for the existence of any possible bound state.
At the bottom quark mass, for both the dibaryons $\Omega_{ccb} \Omega_{ccb}$ and $\Omega_{bbc} \Omega_{ccb}$, $\Delta E$ values found to be far away from their respective NI-energy levels strongly suggesting the possibility of the existence of bound states.
The heavy quark mass dependence in this case therefore displays how an effective binding energy being formed as the quark mass varies from strange to bottom.

\subsection{Heavy quark spin symmetry }
Heavy quark spin symmetry is an approximate symmetry of QCD which suggests that the strong
interactions in the heavy hadrons are independent of the heavy quark
spin. Such a symmetry is found to be useful in studying the decays and production rates of hadrons involving heavy quarks \cite{Neubert:1993mb}. This symmetry is exact in the infinite quark limit and the breaking is of the order of ${\cal{O}}(\Lambda_{QCD}/m_{Q})$, with $m_{Q}$ as the heavy quark mass. 
The dibaryon states in this work provide an opportunity to investigate this symmetry, as we can calculate their masses over a wide range of quark masses, from strange to bottom.
The dibaryon states $\Omega_{bcb} \Omega_{ccb}$ and $\Omega_b \Omega_{bb}$ computed in this work have the exact flavor content as their spin-1 partners, $\mathcal{D}_{bc}$ and $\mathcal{D}_{bs}$, which were already computed and analyzed in previous work in Ref.~\cite{Junnarkar:2019equ}.
They are only different in their spins and will therefore be expected to be degenerate in the heavy quark limit.
Comparing the ground state results of these two spin multiplets will therefore provide access to the effects of heavy quark spin symmetry.
Particularly for dibaryon states $\Omega_{bb}\Omega_b$,  one would expect some breaking of this symmetry since the strange quark is much lighter in mass.
 For the $\Omega_{ccb} \Omega_{bcb}$ dibaryon, since the charm quark is considerably heavier than strange, one should expect a better agreement with its spin-1 partner.
As such, the results are presented in Fig.~\ref{fig:hqs} and we discuss that below. 

In the left panel, the effective masses of deuteron-like spin-1 state, $\mathcal{D}_{bs}$, is represented with the red data and the results of $\Omega_{bb} \Omega_{b}$ with spin-0 is represented by the blue data.
We extract the lowest energy levels  of both of these dibaryons at our finer lattice ensemble ($a=0.0583$ fm) and obtain a mass splitting of:
\begin{equation}
	M^{S=1}_{\Omega_{bb} \Omega_b} - M^{S=0}_{\Omega_{bb} \Omega_b} = 70(30) \ \text{MeV}.
\end{equation}
Such a mass difference can be associated with the size of heavy quark spin symmetry breaking in the dibaryon $\Omega_{bb} \Omega_b$. 
Further, note that the spin-1 state lie above the spin-0 one, a pattern which is consistent with heavy quark hadrons.

The results in  the charm sector are shown in the right panel of Fig.~\ref{fig:hqs}.
With the same color scheme, the spin-1 data of $\mathcal{D}_{bc}$ are shown in red while the spin-0 data of $\Omega_{ccb} \Omega_{bcb}$ is shown in blue.
The level agreement is quite clear and  significantly better than the strange quark counterpart. This is also expected, since the charm quark  is much heavier than strange quark.
Also, the spin-1 effective mass in red data lies slightly above the spin-0. The data are close enough to rule out a clear splitting of the two dibaryons and a fit to the correlators yields,
\begin{equation}
	M^{S=1}_{\Omega_{ccb} \Omega_{bcb}} - M^{S=0}_{\Omega_{ccb} \Omega_{bcb}} = 2(6) \ \text{MeV}.
\end{equation}
This suggests the spin symmetry is exact for these dibaryons. 
Further, this is also manifest in the heavy quark mass dependence of $\Delta E$, which are consistent with those in Ref.~\cite{Junnarkar:2019equ} within the available statistics.

\section{Summary and Discussion\label{sec:disc}}
\begingroup
\renewcommand*{\arraystretch}{1.4}
\begin{table}
\begin{tabular}{c|c} \hline \hline
	Dibaryon &  $\Delta E^{a=0}_{\Omega \Omega}$ \\   \hline \hline
	  $\Omega_{bcb} \Omega_{bcb}$ &  -66(11)\\
	 $\Omega_{ccb} \Omega_{bcb}$ & -48(13)\\
	 $\Omega_{ccb} \Omega_{ccb}$ & -24(16)\\ \hline \hline 
	$\Omega_{bb} \Omega_{bb}$  & -38(40) \\
	 $\Omega_{b} \Omega_{bb}$ & -43(43)\\ \hline \hline
	 $\Omega_{cc} \Omega_{cc}$  & 6(17) \\
	 $\Omega_{c} \Omega_{cc}$ & -16(20)\\ \hline \hline
\end{tabular}
\caption{\label{tab:final_results} Final results of the continuum extrapolated effective binding energies for all the states studied in this work.}
\end{table}
\endgroup
In this work the ground state spectrum of two flavor heavy quark dibaryons in the spin-singlet channel is studied using lattice QCD.
With flavor labels $(l,Q)$ denoting light $l$ and heavy $Q$ flavors, the dibaryon states  $\Omega_{llQ} \Omega_{llQ}, \Omega_{llQ} \Omega_{QlQ}$ and $\Omega_{QlQ} \Omega_{QlQ}$ are investigated.
These states are analogous to spin-singlet states with light quarks that appear in the $NN$ interactions.  In this case, we extend that to two flavor combinations with charm, strange and bottom quarks.
We first consider the physical states with the physical strange, charm and bottom quark masses as shown in Table~\ref{tab:diblist}. Moreover, 
we also investigate how the binding 
energy  depends on the masses  of the constituent quarks by varying the quark masses ($m_l$) over a wide range covering strange to bottom.
The following conclusions are drawn from the analysis of the results:
\begin{enumerate}
	\item In a dibaryon calculation with heavy quarks, the first thing to be determined is the pattern of NI-energy-levels and the corresponding lowest NI-energy-level, as well as the change in the pattern of NI-energy levels with  the change in quark mass.
	 In the bottom sector, the NI-energy-levels from spin-3/2 baryons are found to be the lowest in energy. 
	 In the charm sector, on the other hand, the lowest NI-energy level is found to be that of two spin-1/2 baryons.
	 The lowest energy levels for both cases are shown in the last column in Table~\ref{tab:diblist}. The pattern of their variations are shown in the right panel of Fig.~\ref{fig:NIlevel}. 
	A further interesting finding on the NI-energy levels is that in the bottom sector, for the dibaryons $\Omega_{bcb} \Omega_{bcb}$ and $\Omega_{ccb} \Omega_{ccb}$, the lowest NI-energy levels appear degenerate with no prior expectation for them to be so. 
	\item For each of the spin-singlet dibaryons with botttom and charm quarks, namely for $\Omega_{bcb} \Omega_{bcb}$,  $\Omega_{ccb} \Omega_{bcb}$ and  $\Omega_{ccb} \Omega_{ccb}$, we find an energy level below their respective lowest NI-energy-level. The large energy  difference ($\Delta E$) between their ground states and respective lowest NI-energy levels indicates the possibility of the existence of deeply bound dibaryons for these cases. Since  $\Delta E$ values are large and these dibaryons are solely made of heavy quarks, it is unlikely that there will be substantial volume effect to shift their ground state energies towards their NI-energy levels making them unbound. Nevertheless, this needs to be confirmed with a rigorous finite volume amplitude analysis.  
 
  The results in the bottom-strange sector have larger uncertainties and prohibit a clear identification of a bound state. In the charm-strange sector, the ground state energy levels, within the limited statistics used in this work, are found to be consistent with the relevant NI-energy levels,  and hence do not  provide an indication of any bound state. Absence of a possible bound state for this case becomes  more clearer when we replace the strange quark by the charm quark and still find $\Delta E$ to be consistent with zero, even on our finer ensemble.
Our final results of $\Delta E$ with continuum extrapolation are shown in Table~\ref{tab:final_results}. For the cases of bottom sector we use ensembles with three different lattice spacings and for the charm sector ensembles with four lattice spacings are employed.
	 \item Our study indicates an intriguing feature of strong interactions:-- heavier the quark masses stronger is the binding in dibaryons. We make this observation by investigating the quark mass dependence of possible binding of dibaryons $\Omega_{QlQ} \Omega_{QlQ}$ and  $\Omega_{llQ} \Omega_{QlQ}$, and varying the quark mass $m_l$ over a wide range, $m_s \le m_l \le m_b$, while keeping $m_Q$ at the physical charm or bottom quark mass.
The trend in the possible binding energies ($\Delta E$) in the charm sector shows that at $m_l = m_s$ and $m_l = m_c$, $\Delta E$ is consistent with zero. However, beyond the charm quark mass,  as the flavor $l$ gets heavier towards the bottom ($m_c \le m_l \le m_b$), there is a clear trend that $\Delta E$ is non-zero and becoming deeper as  $m_l$ increases towards $m_b$ suggesting the formation of bound states. In the bottom sector, as the flavor $l$ gets heavier, $\Delta E$ becomes even larger, suggesting the existence of bound states with even deeper bindings.  Similar pattern of quark mass dependence of possible binding energies of spin-1 dibaryons and three flavor H-type heavy dibaryons were found earlier in Ref.~\cite{Junnarkar:2019equ,Junnarkar:2022yak}. Together these results clearly suggest that
heavier the quark masses stronger is the binding in dibaryons. Perhaps, origin of such a deep binding is different than that of our phenomenological understanding of the NN interactions through pion-exchange. The physical sizes of these dibaryons could be very small and they could be tightly bound by one or multigluon exchanges. However, a proper effective field theory explanation of the origin of this deep binding is needed and we hope that our works will motivate such studies.

	 \item Another interesting aspect of this work is to provide an opportunity 
  to study the heavy quark spin symmetry of heavy hadrons by comparing ground state masses of spin-0 and spin-1 heavy dibaryons. In particular, the ground state masses of  spin-0 states, $\Omega_{ccb} \Omega_{bcb}$, are compared to corresponding spin-1 states, $\mathcal{D}_{bc}$, from Ref.~\cite{Junnarkar:2019equ}, and similarly $\Omega_{b} \Omega_{bb}$ to $\mathcal{D}_{bs}$. These dibaryons have the same quark contents and they differ only in their spin quantum numbers.
		The comparison of their ground state masses and their differences are shown in Figure~\ref{fig:hqs}.
	In the presence of a strange quark in each of the baryons in $\mathcal{D}_{bs}$ and $\Omega_b \Omega_{bb}$, there is a splitting of about 70(30) MeV. This is similar order of mass splitting found between $\Omega^{*}_b (3/2)$ and  $\Omega_b (1/2)$ baryons~\cite{Mathur:2018rwu,Brown:2014ena}.  In the bottom sector, with $\mathcal{D}_{bc}$ (spin-1) and $\Omega_{bcb}\Omega_{ccb}$ (spin-0) dibaryons, the observed splitting is consistent with zero suggesting the presence of near-exact spin symmetry at this limit. This is also consistent with the mass splitting between $\Omega^*_{cbb} (3/2)$ and $\Omega^*_{cbb} (1/2)$ baryons~\cite{Mathur:2018epb}.
\end{enumerate}

Given these results, suggesting the possibility of the existence of a number of deeply bound heavy dibaryon states, it would be essential to carry out a detail finite volume study in future investigating their pole structures which will help to make a firm conclusion about these states. Although the search for these possible bound dibaryons is not feasible at the moment, it could well become feasible if multiple triply-heavy baryons are discovered in future as and when higher luminosity at the required center-of-momentum energy  becomes available at high energy laboratories, particularly at LHCb. 
Moreover, our results also suggest that heavier the quark masses, stronger is the binding in a dibaryon.  A detailed theoretical understanding of that, using effective field theories and phenomenological models, will be highly sought after, as it may provide insights into how binding emerges in a dibaryon as the mass of a quark changes.

\begin{acknowledgments}
  This work is supported by the Department of Atomic Energy, Government of India, under Project Identification Number RTI 4002. PMJ would like to acknowledge support by the Deutsche
Forschungsgemeinschaft (DFG, German Research Foundation) through the CRC-TR 211 ``Strong-interaction matter under extreme conditions''- project number
315477589 - TRR 211. PMJ thanks the Institut f{\"u}r Kernphysik at TU-Darmstadt where most of the work was done and  the Research Center for Nuclear Physics at Osaka University and the Wako Kiban (S) collaboration for supporting this research. 
We are thankful to the MILC collaboration and in particular to S. Gottlieb for providing us with the HISQ lattices. Computations are carried out on the Cray-XC30 of ILGTI, TIFR,   and on the Pride/Flock clusters of the Department of Theoretical Physics,
TIFR.   NM would also like to thank Ajay Salve, Kapil Ghadiali and T. Chandramohan for computational support. 
\end{acknowledgments}

\bibliographystyle{utphys-noitalics}
\bibliography{2fbiblio}
\end{document}